\input ./style/arxiv-general.cfg
\documentclass[aoas,MSNbibl,nameyear,rotating,seceqn,dvips]{arximspdf}
\makeatletter
   \@ifpackageloaded{graphicx}{}{\usepackage{graphicx}}
\makeatother
\usepackage{accents}
\usepackage{dcolumn}
\usepackage{multirow}

\doi{10.1214/15-AOAS826}
\volume{9}
\issue{2}
\pubyear{2015}
\firstpage{731}
\lastpage{753}
\docsubty{FLA}

\makeatletter
\newcolumntype{d}[1]{D{.}{.}{#1}}
\newcommand{\rrvert}{\vert}
\newcommand{\llvert}{\vert}
\makeatother

\begin{document}
\begin{frontmatter}

\title{Biased sampling designs to improve research efficiency: Factors
influencing pulmonary function~over time in children with
asthma\thanksref{T1}}
\thankstext{T1}{Supported in part  by the National Heart Lung and
Blood Institute Grants R01 HL094786 and R01 HL072966.}
\runtitle{Biased sampling designs to improve research efficiency}

\begin{aug}
\author[A]{\fnms{Jonathan S.}~\snm{Schildcrout}\corref{}\thanksref{M1}\ead[label=e1]{jonathan.schildcrout@vanderbilt.edu}},
\author[B]{\fnms{Paul~J.}~\snm{Rathouz}\thanksref{M2}\ead[label=e2]{rathouz@biostat.wisc.edu}},
\author[C]{\fnms{Leila~R.}~\snm{Zelnick}\thanksref{M3}\ead[label=e3]{zelnick@u.washington.edu}},
\author[D]{\fnms{Shawn~P.}~\snm{Garbett}\thanksref{M1}\ead[label=e4]{shawn.garbett@vanderbilt.edu}}
\and
\author[C]{\fnms{Patrick J.}~\snm{Heagerty}\thanksref{M3}\ead[label=e5]{heagerty@u.washington.edu}}
\runauthor{J. S. Schildcrout et al.}
\affiliation{Vanderbilt University School of Medicine \thanksmark{M1}, University of
Wisconsin\\ School of Medicine and Public Health\thanksmark{M2} and University of\\ Washington School of Public~Health\thanksmark{M3}}
\address[A]{J. S. Schildcrout\\
Departments of Biostatistics\\
\quad  and Anesthesiology \\
Vanderbilt University School of Medicine \\
2525 West End Ave, Suite 11000\\
Nashville, Tennessee 37203\\
USA\\
\printead{e1}}
\address[B]{P. J. Rathouz \\
Department of Biostatistics\\
\quad and Medical Informatics \\
University of Wisconsin School of Medicine\\
\quad and Public Health\\
K6/446 Clinical Sciences Center\\
600 Highland Avenue\\
Madison, Wisconsin 53792-4675\\
USA\\
\printead{e2}}
\address[C]{L. R. Zelnick \\
P. J. Heagerty \\
Department of Biostatistics \\
University of Washington \\
Box 357232\\
1959 NE Pacific Street\\
Seattle, Washington 98195\\
USA\\
\printead{e3}\\
\phantom{E-mail:\ }\printead*{e5}}
\address[D]{S. P. Garbett\\
Department of Cancer Biology\\
Vanderbilt University School of Medicine\hspace*{7pt}\\
2220 Pierce Ave\\
Nashville, Tennessee 37232\\
USA\\
\printead{e4}}
\end{aug}

%
\received{\smonth{12} \syear{2013}}
%
\revised{\smonth{3} \syear{2015}}

\begin{abstract}
Substudies of the Childhood Asthma Management Program [\textit{Control. Clin. Trials} \textbf{20} (1999)
91--120; \textit{N. Engl. J. Med.} \textbf{343} (2000) 1054--1063] seek to identify patient characteristics
associated with asthma symptoms and lung function. To determine
if genetic measures are associated with trajectories of lung function
as measured by forced vital capacity (FVC), children in the primary
cohort study retrospectively had candidate loci evaluated. Given participant
burden and constraints on financial resources, it is often desirable
to target a subsample for ascertainment of costly measures.
Methods that can leverage the longitudinal outcome on the full cohort
to selectively measure informative individuals have been promising,
but have been restricted in their use to analysis of the targeted
subsample. In this paper we detail two multiple imputation analysis
strategies that exploit outcome and partially observed covariate data
on the nonsampled subjects, and we characterize alternative design
and analysis combinations that could be used for future studies of pulmonary
function and other outcomes. Candidate predictor (e.g., IL10
cytokine polymorphisms) associations obtained from targeted sampling
designs can be estimated with very high efficiency compared
to standard designs. Further, even though multiple imputation can
dramatically improve estimation efficiency for covariates available on
all subjects (e.g., gender and baseline age), relatively modest efficiency
gains were observed in parameters associated with predictors that
are exclusive to the targeted sample. Our results suggest that future
studies of longitudinal trajectories can be efficiently conducted by use
of outcome-dependent designs and associated full cohort analysis.
\end{abstract}

\end{frontmatter}

\section{Introduction}\label{sec1}

The Childhood Asthma Management Program [CAMP; CAMP Research Group
(\citeyear{pmid10027502} and \citeyear{CAMP2000})] was a randomized clinical trial that compared two
anti-inflammatory medications and a placebo on lung growth over the
course of 4 years
in children with mild to moderate asthma.
CAMP substudies have since examined the relationship between genetic
factors and asthma phenotypes. Like other genetic data collected in
CAMP, interleukin-10 (IL10) genotype data were obtained retrospectively
by analysis of stored blood samples. IL10 is a type-2 T-helper cytokine
with anti-inflammatory properties, and polymorphisms in the IL10 cytokine
gene have been shown to be associated with asthma phenotypes in children
[Lyon et al. (\citeyear{L04})]. However, as is often the case, ascertainment of
expensive exposures can restrict sample size and therefore motivate
thoughtful sampling strategies. Given that the outcome of interest was
available on all subjects, we seek to determine whether
the longitudinal response could or should be used to target a subset of
select individuals for sampling of covariates.
In particular, we explore both sampling designs and associated analysis
options with the goal of providing recommendations for the efficient conduct
of future retrospective studies.

We are specifically interested in the impact genetic variants have on
both lung function and growth, and on the effect of medication
(versus placebo) within subgroups defined by genetic variants of the
IL10 gene.
For nearly all children, forced vital capacity (FVC, a measure of lung
function) was measured ten times over the course of 4 years, thereby
providing rich detail on the primary response trajectory.
Our scientific question can be addressed by appropriate
longitudinal regression models with a focus on
estimating the main effects of time since randomization,
time-invariant randomized treatment assignment
(Budesonide, Nedocromil, placebo), and their
interactions with the presence or absence of at least one
IL10 polymorphism.
Valid IL10 and other data were available for 555 children who
participated in CAMP. Even though all data were available for these
children, we will illustrate the interplay between sampling strategies
and analysis procedures by assuming study resources are limited and
IL10 data can only be collected on approximately 250 children. The
assumption of limited resources allows us to compare and contrast several
sampling designs and estimation procedures in order to inform decisions
when conducting similar substudies in the future.

In related work, Neuhaus, Scott and Wild (\citeyear{NeuhScotWildanal2002}, \citeyear{NeuhScotWildfami2006}) discussed biased,
outcome dependent sampling (ODS) designs with longitudinal response
data and estimation from resulting data using a profile likelihood. In
the longitudinal binary response setting, {Schildcrout and Heagerty
(\citeyear{pmid18372397}, \citeyear{pmid21457191})} described stratified sampling designs based on the sum of
the response series with an ascertainment \mbox{corrected} likelihood approach
for analysis. \citet{pmid19673861}, Schildcrout et al. (\citeyear{pmid22086716}) and
Neuhaus et al. (\citeyear{pmid24571396}) addressed auxiliary variable
dependent sampling where the sampling variable is related but is not
equal to the longitudinal response. In the univariate continuous
response setting, Zhou et al. (\citeyear{ZhouWeavQinLongWangsemi2002}, \citeyear{pmid17568219}) and \citet{WeavZhouan2005}
discussed ODS designs that combine simple random samples with a sample
of subjects whose responses are more extreme. Further, several authors
discussed unplanned outcome-dependent follow-up for longitudinal
continuous response data [e.g., \citet{ISI000167326000010,pmid12229997,pmid19035546}].

In \citet{pmid23409789}, we proposed biased epidemiological study
designs for continuous longitudinal response data where sampling is
based on strata defined by low-dimensional summaries of the response
series. We proposed sampling based on the intercept, the slope, or both
the intercept and slope of the subject-specific ordinary least squares
(OLS) regressions of the response on a time-varying covariate (which
may be time itself). We showed that sampling based on a variable
related to a target predictor can lead to substantial efficiency gains
relative to random sampling for the associated parameter. Such a result
is well known to survey sampling methodologists [e.g., see \citet{kish1965survey,korn2011analysis}]. The estimation procedure discussed in
\citet{pmid23409789} used a bias correcting, ascertainment
corrected conditional likelihood that only includes subjects with fully
observed exposure data (i.e., those who were sampled). Such an analysis
can be referred to as a complete data (CD) analysis
[Carroll et al. (\citeyear{carroll2006measurement}), \citet{ISI000079074800009}]. In univariate response settings, such
as the case-cohort design, other authors [e.g., Breslow et al. (\citeyear{pmid20174455,pmid19357328}), \citet{pmid21351290}] have shown that utilizing the partial data
on the unsampled subjects can add information and improve estimation efficiency.

With specific motivation from the CAMP study, the purpose of this manuscript
is to detail the joint impact of sampling design and statistical analysis
decisions toward efficient parameter estimation with longitudinal
continuous response data.
Longitudinal outcome-dependent sampling designs have only recently been
proposed, and analysis options have not considered use of both sampled and
unsampled subjects.
Using the CAMP study for motivation and illustration, we focus on the following
goals: (1) to evaluate circumstances under which multiple imputation (MI)
increases efficiency appreciably over the bias-correcting complete data (CD)
analysis under ODS designs, and (2) to evaluate the extent to which the ODS
designs improve estimation efficiency when MI (rather than CD analysis) is
the chosen analytical approach.
We use a simulation study to explore relative efficiency across several
sampling design and estimation procedure combinations.
The CAMP study is an exemplar of a longitudinal randomized
trial in which retrospective collection of additional explanatory data is
conducted to in order to leverage the original cohort study and answer new
scientific questions. The CAMP data provide an ideal context to inform
efficient study design options for future ancillary studies of
factors associated with longitudinal outcome trajectories.

Section~\ref{Model} discusses the model of interest, briefly reviews
the sampling strategy and estimation procedure discussed in
\citet{pmid23409789}, and proposes two multiple imputation analysis
strategies that exploit the unsampled subjects' data. Section~\ref{Simulation} examines the relative efficiency of design and analysis
procedures in a number of plausible scenarios. Section~\ref{CAMP}
returns to the CAMP data to examine the impact of study designs on the
FVC data, and Section~\ref{Discussion} provides a discussion including
directions for future research.

\section{Methodological framework}
\label{Model}

We now introduce the mixed model, the class of ODS designs and
associated CD analyses, 
and two multiple imputation (MI) extensions for conducting analyses.

\subsection{Linear mixed effects model for continuous longitudinal
response data}

With $N$ subjects in the original cohort, $\mathbf{Y}_i$, $i \in1 ,2,
\ldots,
N$, the $n_i$-vector of response values, $\mathbf{X}_i$, a $n_i \times p$
fixed effects design matrix, and $\mathbf{Z}_i$ the $n_i \times q$ design
matrix for the random effects, we begin with the \citet{pmid7168798}
linear mixed effects model given by
%
\begin{equation}
\mathbf{Y}_i= \mathbf{X}_i\bolds{\beta}+
\mathbf{Z}_i\mathbf{b}_i + \bolds{\varepsilon}_i,
\end{equation}
where $\bolds{\beta}$ is a $p$-vector of fixed-effect
coefficients, $\mathbf{b}_i \sim N(\mathbf{0}, \mathbf{D})$, and
$\bolds{\varepsilon}_i \sim N(0, \bolds{\Sigma})$. A common design
matrix for the random
effects in the continuous data setting is $\mathbf{Z}_i= (\mathbf{1},
\mathbf{T}_i)$, where $\mathbf{T}_i$ is a time-varying
covariate---perhaps time
itself, $\mathbf{b}_i = (b_{0i}, b_{1i})$, and $\mathbf{D}_i$ is the $2
\times2$ covariance matrix containing variance components $(\sigma
_0^2, \sigma_1^2)$ and correlation $\rho=\operatorname{corr}(b_{0i}, b_{1i})$.
Analysis based on a random sample of $N_s$ subjects can be conducted by
maximizing the log-likelihood
%
\begin{equation}
\label{LairdWare}
l(\bolds{\theta}; \mathbf{Y}, \mathbf{X}) = \sum
_{i=1}^{N_s} l_i(\bolds{\theta };
\mathbf{Y}_i, \mathbf{X}_i) = \sum
_{i=1}^{N_s} \log f(\mathbf{Y}_i\vert
\mathbf{X}_i; \bolds{\theta}),
\end{equation}
where $\bolds{\theta}=(\bolds{\beta}, \sigma_0, \sigma_1, \rho
)$ and $f(\cdot)$
is the multivariate normal density function.

\subsection{Coarsened summary sampling designs} \label{Design}

Study designs proposed in  \citet{pmid23409789} propose
subsampling from a larger cohort based on a user defined,
low-dimensional summary of the outcome vector $\mathbf{Y}_i$ or, more
accurately, on strata
defined by the summary measure. Let $\mathbf{X}_{oi}$ be a covariate
subset of $\mathbf{X}_i$ that is known prior to initiation of the substudy
and let $\mathbf{Q}_i= g(\mathbf{Y}_i, \mathbf{X}_{oi})$ be any
function of the
response and observed covariates that summarizes important features of
the response vectors. Three simple and useful summaries are the
estimated intercept, slope, and the joint intercept and slope, based on
the subject-specific OLS regression of 
$\mathbf{Y}_i$ on a time-varying covariate. For example, if $\mathbf{T}_{i}$
is the easily ascertained time-varying covariate, $\mathbf{X}_{ti} =
(\mathbf{1}, \mathbf{T}_{i}) \subset\mathbf{X}_{oi}$, and $\mathbf
{W}_{oi}=(\mathbf{X}_{ti}^t \mathbf{X}_{ti} )^{-1}\mathbf
{X}_{ti}^t$, then $\mathbf{Q}_i=\mathbf{W}_{oi} \mathbf{Y}_i$ is the
estimated intercept and slope for the
regression of $\mathbf{Y}_i$ on $\mathbf{T}_{i}$. We proposed stratified
random sampling based on regions of $\mathbf{Q}_i$. Based on results from
other literature [e.g., Zhou et al.
(\citeyear{ZhouWeavQinLongWangsemi2002,pmid17568219,pmid21252082})], we
oversampled the extremes of the $\mathbf{Q}_i$ distribution to realize
substantial efficiency gains for target parameters. Let $S_i$ equal $1$
if subject $i$ is sampled for exposure ascertainment and 0 if not. For
region $R^k \in\{R^1, \dots, R^K\}$, let $\pi(R^k)= \operatorname{pr}(S_i =
1 \vert \mathbf{Y}_i, \mathbf{X}_i) = \operatorname{pr}(S_i =1 \vert \mathbf
{q}_i \in
R^k)$ be the probability of being sampled given $\mathbf{q}_i$, the
observed value of $\mathbf{Q}_i$, is in region~$k$. Importantly, $S_i
\perp(\mathbf{Y}_i, \mathbf{X}_i) \vert \mathbf{q}_i$, that is,
sampling depends
upon the data $(\mathbf{Y}_i, \mathbf{X}_i)$ only through $\mathbf{q}_i$.

\subsection{An ascertainment corrected likelihood for coarsened
summary sampling designs} \label{ACL}

For inferences to the population represented by the original
cohort---as opposed to the pseudo-population represented by the biased
sample---\citet{pmid23409789} considered maximization of an
ascertainment corrected likelihood (ACL). The ACL corrects for the
design by conditioning the likelihood on inclusion into the ODS
($S_i=1$). It is a ``complete data'' (CD) likelihood
[Carroll et al. (\citeyear{carroll2006measurement}), \citet{ISI000079074800009}] in that only subjects with complete
exposure data contribute to the conditional likelihood, and therefore
to the analysis. A key attraction of the CD approach is that valid
inferences can be realized while only requiring a model for $\mathbf
{Y}_i\vert
\mathbf{X}_i$ without requiring a model for $\mathbf{X}_i$.
Specifically, if
$f(\mathbf{Y}_i\vert \mathbf{X}_i; \bolds{\theta}) $ is the density
for subject $i$
under simple random sampling from a population, the density for those
who are included in the ODS is given by
%
\begin{eqnarray}
&& f(\mathbf{Y}_i\vert \mathbf{X}_i, S_i=1 ;
\bolds{\theta})\nonumber\\
\label{ConditionalDensity}
 &&\qquad = \pi(\mathbf{q}_i ) f(\mathbf{Y}_i
\vert \mathbf{X}_i; \bolds{\theta }) \bigl\{\operatorname{pr}(S_i=1
\vert \mathbf{X}_i; \bolds{\theta}) \bigr\}^{-1}
\\
\nonumber
&&\qquad = \pi(\mathbf{q}_i) f(\mathbf{Y}_i\vert
\mathbf{X}_i; \bolds{\theta}) \Biggl\{\sum
_{k=1}^K \pi\bigl(R^k\bigr) \int
_{R^k} f(\mathbf{q}_i \vert \mathbf{X}_i;
\bolds{\theta })\,d\mathbf{q}_i \Biggr\} ^{-1},
\end{eqnarray}
where
$\pi(\mathbf{q}_i)$ is subject $i$'s sampling probability that is based
on $\mathbf{q}_i$ [i.e., $\pi(\mathbf{q}_i) = \pi(R^k)$ if and only
if $\mathbf{q}_i \in R^k$], $\pi(R^k)$ is the sampling probability
for all values
of $\mathbf{Q}_i$ in region $R^k$, and $\int_{R^k} f(\mathbf{q}_i
\vert \mathbf{X}_i
; \bolds{\theta})\,d\mathbf{q}_i =\operatorname{pr}(\mathbf{q}_i \in R^k
\vert \mathbf{X}_i; \bolds{\theta})$. Because $\pi(\mathbf{q}_i)$
is parameter-free,
being specified by the investigator, if a total of $N_s$ subjects are
selected into the ODS for exposure ascertainment, the ascertainment
corrected log-likelihood, $l^C(\bolds{\theta};\mathbf{Y}, \mathbf
{X})$, is
given by
%
\begin{eqnarray}\label{loglike}
&& l(\bolds{\theta}; \mathbf{Y}, \mathbf{X}) - \sum_{i=1}^{N_s}
\log \Biggl\{ \sum_{k=1}^K \pi
\bigl(R^k\bigr) \int_{R^k} f(\mathbf{q}_i
\vert \mathbf{X}_i; \bolds{\theta})\,d\mathbf{q}_i \Biggr
\}.
\end{eqnarray}

In the special case where $\mathbf{Q}_i = \mathbf{W}_{oi} \mathbf
{Y}_i$ is a
linear transformation of $\mathbf{Y}_i$, under the assumption $\mathbf{Y}_i
\vert \mathbf{X}_i \sim N(\bolds{\mu}_i, \mathbf{V}_i)$, then
$\mathbf{Q}_i \vert
\mathbf{X}_i \sim N(\bolds{\mu}_{q,i}, \mathbf{V}_{q,i})$, where
$\bolds{\mu }_{q,i}=\mathbf{W}_{oi} \bolds{\mu}_i$ and $\mathbf
{V}_{q,i} = \mathbf{W}_{oi}
\mathbf{V}_i \mathbf{W}_{oi}^t$. Thus, the ACL is a straightforward extension
of the likelihood used for standard analyses, and details can be found
in \citet{pmid23409789}. We note that this log-likelihood is
composed of two terms: the standard log-likelihood as in equation (\ref
{LairdWare}) and an additive ascertainment correction piece that
accounts for the biased study design and is the probability of being
sampled as a function of~$\mathbf{X}_{oi}$. This is in contrast to inverse
probability weighting or weighted likelihood approaches
[e.g., Horvitz and Thompson (\citeyear{H52}), \citet{RobiRotnZhaoesti1994}] that multiply the
log-likelihood by a function of the sampling probability to calculate
an unbiased estimating equation.

\subsection{Multiple imputation} \label{MISection}

Whereas the analysis procedures proposed in \citet{pmid23409789}
were more efficient than random sampling, one can expect that there may
be additional information in those subjects for whom the unmeasured,
expensive exposure, $X_{ei}$, was not ascertained (i.e., those with
$S_i=0$). We therefore propose to multiply impute [\citet{ISIA1976CP66700021}]
$X_{ei}$ for all subjects in whom $S_i=0$. Multiple imputation (MI) is
expected to recover some of the information about the parameter
associated with $X_{ei}$ that is lost by not measuring $X_{ei}$, and it
is expected to recover much more of the information in parameters
associated with $\mathbf{X}_{oi}$ that is available but is not used in CD
analyses. Multiple imputation is attractive because it can leverage
existing methods and software without needing tailored programs. In the
approaches described below, we generate imputation samples from the
conditional exposure distribution in unsampled subjects $[X_{ei} \vert
\mathbf{Y}_i, \mathbf{X}_{oi}, S_i=0]$. Once the exposure model is
constructed, we build $M$ multiple imputation data sets, fit the target
model to each one using standard maximum likelihood, and combine
estimates across imputations to make inferences regarding model
parameters. For any\vspace*{1.5pt} parameter $\theta$ in $\bolds{\theta}$, we may
estimate its value and\vspace*{1.5pt} variance with $\widehat{\theta} = M^{-1} \sum_{m=1}^M \widehat{\theta}^{(m)}$ and
$\widehat{\operatorname{Var}}(\widehat{\theta})=\overline{V} + (1+M^{-1})B$, respectively,
where $\overline{V} = M^{-1} \sum_{m=1}^M \widehat{\operatorname{Var}}(\widehat{\theta}^{(m)})$ and $B=(M-1)^{-1} \sum_{m=1}^M
(\widehat{\theta}^{(m)} - \widehat{\theta})^2$. With adequate $M$,
test statistics for parameters are well approximated by a\vadjust{\goodbreak} standard
Gaussian distribution; however, with small $M$, a $t$-distribution with
$\mathit{df}=(M-1)[1+M \overline{V} / \{(M+1)B\} ]^2$ degrees of freedom is
more appropriate [\citet{ISIA1976CP66700021,LittleRubin2002,SchaferGraham2002}]. In the settings we believe our designs could be most
useful, $X_{ei}$ is to be imputed in a relatively large percentage of
subjects (i.e., well over 50 percent), and in such cases a larger
number of imputation samples are required to use the normal
approximation to the $t$-distribution.

We now describe two approaches to estimating the imputation model
$[X_{ei} \vert \mathbf{y}_i, \mathbf{x}_{oi}, S_i=0]$. The first is an
extension of the CD analysis described in Section~\ref{ACL} and the
second is a direct imputation approach that does not require estimation
based on maximizing the ACL. Because the ODS sampling schemes we have
described depend upon the data through a low-dimensional response
summary and possibly observed covariates $\mathbf{X}_{oi}$,
\begin{equation}\label{MIModel}
\operatorname{pr}(x_{ei} \vert \mathbf{x}_{oi},
\mathbf{y}_i, S_i=0) = \operatorname{pr}(x_{ei}
\vert \mathbf{x}_{oi}, \mathbf{y}_i) = \operatorname{pr}(x_{ei}
\vert \mathbf{x}_{oi}, \mathbf{y}_i, S_i=1).
\end{equation}
%
Thus, the design-based ``missing data mechanism'' is ignorable and
generating $X_{ei}$ for unsampled subjects can be based directly on
model estimates derived from sampled data without consideration of the
biased sample. Importantly, for the CAMP analysis, the missing exposure
variable ($X_{ei}$) was binary and so for the present research, we only
detail this special case explicitly; however, extensions to continuous
and other exposure values are feasible.

\subsubsection{Imputation model construction: Combine response model
and mar\-ginal exposure model} \label{MIApproachesCDMI}

The complete data plus multiple imputation analysis approach (CD$+$MI)
combines the estimates from maximizing the ACL in Section~\ref{ACL}
with an exposure model for $[X_{ei} \vert \mathbf{x}_{oi}, S_i=1]$ to
estimate $[X_{ei} \vert \mathbf{y}_i,\break \mathbf{x}_{oi}, S_i=0]$. Specifically,
we combine a CD estimate of $[\mathbf{Y}_i\vert \mathbf{x}_i, S_i=1]$
with a
covariate logistic regression for $[X_{ei} \vert \mathbf{x}_{oi}, S_i=1]$
to identify the conditional exposure distribution $[X_{ei} \vert \mathbf
{y}_i, \mathbf{x}_{oi}, S_i=1]$ used for imputation among those
with~$S_i=0$. Using equation~(\ref{MIModel}) and\vspace*{-2pt} Bayes' theorem,
%
\begin{eqnarray}
&&
\frac{\operatorname{pr}(X_{ei}=1 \vert \mathbf{x}_{oi}, \mathbf{y}_i,
S_i=0)}{\operatorname{pr}(X_{ei}=0 \vert \mathbf{x}_{oi}, \mathbf{y}_i, S_i=0)}
\nonumber
\\[-8pt]
\label{MIodds}\\[-8pt]
\nonumber
&&\qquad=  \frac{f(\mathbf{y}_{i} \vert  X_{ei}=1, \mathbf{x}_{oi},
S_i=1)}{f(\mathbf{y}_{i}
\vert  X_{ei}=0, \mathbf{x}_{oi}, S_i=1)} \cdot \frac{\operatorname{pr}(X_{ei}=1 \vert \mathbf{x}_{oi}, S_i=1)}{\operatorname{pr}(X_{ei}=0 \vert \mathbf{x}_{oi}, S_i=1)}.
\end{eqnarray}
Using the logistic regression model to obtain estimate $\widehat
{\operatorname{pr}}(x_{ei} \vert \mathbf{x}_{oi}, S_{i}=1)$ in the observed subjects'
data, and then combining it with $\widehat{f}(\mathbf{y}_{i} \vert  x_{ei},
\mathbf{x}_{oi}, S_i=1)$ from the CD analysis, we are able to estimate and sample
from $\widehat{\operatorname{pr}}(x_{ei} \vert \mathbf{x}_{oi}, \mathbf{y}_i,\break S_i=0)$.

\paragraph*{Note} We may write the exposure odds model itself as
\begin{eqnarray}
&& \frac{\operatorname{pr}(X_{ei}=1 \vert \mathbf{x}_{oi}, S_i=1)}{\operatorname{pr}(X_{ei}=0 \vert \mathbf{x}_{oi}, S_i=1)}
\nonumber
\\[-8pt]
\label{Exposureodds}
\\[-8pt]
\nonumber
&&\qquad = \frac{\operatorname{pr}(S_i=1 \vert  X_{ei}=1, \mathbf{x}_{oi})}{\operatorname{pr}(S_i=1 \vert  X_{ei}=0, \mathbf{x}_{oi})} \cdot \frac{\operatorname{pr}(X_{ei}=1 \vert \mathbf{x}_{oi})}{\operatorname{pr}(X_{ei}=0
\vert \mathbf{x}_{oi})}.
\end{eqnarray}
The first term on the right side of the equation is a ratio of the
ascertainment corrections for $X_{ei}=1$ and $X_{ei}=0$ that is shown
in equation (\ref{ConditionalDensity}). We can therefore use the log
of the ratio of ascertainment corrections as an offset in a logistic
regression, marginal exposure model given by (\ref{Exposureodds}). In
some cases, such an approach may be more natural or simple than
modeling the marginal exposure model on the left side of equation (\ref
{Exposureodds}) directly. This is due to the fact that the marginal
exposure model, $\operatorname{pr}(X_{ei} \vert \mathbf{X}_{oi})$, may be
simpler in the population as compared to the observed sample, $\operatorname{pr}(X_{ei} \vert \mathbf{X}_{oi}, S_i=1)$. For example, in many realistic
scenarios, one would expect that time-varying and time-invariant
covariates are independent in the population. In the CAMP, time since
randomization is expected to be independent of, say, genotype. However,
for the biased sample, such time-varying covariates may be spuriously
associated with genotype due to their impact on the probability of
being sampled. If one wished to model the left-hand side of equation
(\ref{Exposureodds}) directly, the functional forms of time-varying
covariates must be carefully considered.

The steps for creating the imputation data sets used in the CD$+$MI
approach are as follows:
\begin{enumerate}[(3)]
\item[(1)] On\vspace*{1,5pt} sampled subjects, $S_i=1$, maximize the ascertainment
corrected log-likelihood\vspace*{1pt} shown in equation (\ref{loglike}) to obtain
estimates $\widehat{\bolds{\theta}}$ and uncertainty $\widehat
{\operatorname{Cov}}(\widehat{\bolds{\theta}})$ associated with the response model.
\item[(2)] For\vspace*{1pt} $m=1,\ldots,M$, draw $\bolds{\theta}^{(m)}$ from the
approximate posterior distribution for $\widehat{\bolds{\theta}}$ given
by the normalized likelihood function, and calculate
\begin{enumerate}[(2a)]
\item[(2a)] \vspace*{2pt}$f(\mathbf{y}_{i} \vert  X_{ei}=1, \mathbf{x}_{oi}, S_i=1;
\bolds{\theta}^{(m)})  \{ f(\mathbf{y}_{i} \vert  X_{ei}=0,
\mathbf{x}_{oi},
S_i=1; \bolds{\theta}^{(m)})  \}^{-1}$,
\item[(2b)] $\log [ \operatorname{pr}(S_i=1 \vert  X_{ei}=1, \mathbf
{x}_{oi}; \bolds{\theta}^{(m)})  \{ \operatorname{pr}(S_i=1 \vert
X_{ei}=0, \mathbf{x}_{oi}; \bolds{\theta}^{(m)}  \}^{-1}
 ]$.
\end{enumerate}
\item[(3)] On sampled subjects, using (2b) as an offset, fit a
logistic regression of $X_{ei}$ on $\mathbf{X}_{oi}$ to obtain parameter
(call it $\bolds{\alpha}$) and uncertainty estimates for the marginal
exposure model shown in equation (\ref{Exposureodds}). Then, draw
$\bolds{\alpha}^{(m)}$ from a $N[\widehat{\bolds{\alpha}},
\widehat{\operatorname{Cov}}(\widehat{\bolds{\alpha}})]$ and calculate
\begin{enumerate}[(3a)]
\item[(3a)] $\operatorname{pr}(X_{ei}=1 \vert \mathbf{x}_{oi}, S_i=1;
\bolds{\alpha}^{(m)})  \{ \operatorname{pr}(X_{ei}=0 \vert
\mathbf{x}_{oi}, S_i=1; \bolds{\alpha}^{(m)})  \}^{-1}$.
\end{enumerate}
\item[(4)] For unsampled subjects, multiply the results of (2a) and
(3a) to calculate the conditional exposure odds in equation (\ref
{MIodds}) and then draw imputed values, $X_{ei}^{(m)}$.
\item[(5)] Conduct standard maximum likelihood analysis on the
response model using the complete imputation data set.
\item[(6)] Repeat steps (2)--(5) $M$ times and combine results in
the standard manner.
\end{enumerate}
To the extent that the assumptions of the response and marginal
exposure models are correct, the foregoing CD$+$MI approach is expected
to be valid and relatively efficient compared to the CD approach. It is
worth noting that the imputation model for the CD$+$MI approach is a
general location model that is discussed in, for example,\vspace*{-2pt} \citet{little1985maximum,schafer2010analysis}, and \citet{LittleRubin2002}.

\subsubsection{Imputation model construction: Direct conditional
exposure model} \label{MIApproachesDMI}

Another approach to constructing the imputation model is relatively
direct and could employ available MI software. In contrast to CD$+$MI, it
decouples the imputation and the analysis models. We refer to it as
direct multiple imputation (D-MI) and it is a special case of multiple
imputation by chained equations [e.g., \citet{Raghunathan++01,ISI000287106200008}] which is implemented in software packages such as MICE
[\citet{van2012flexible}] in the R programming language [\citet{R}].
We may ascertain and sample from $[X_{ei} \vert \mathbf{y}_i, \mathbf{x}_{oi},
S_{i}=0]$ directly by noting that the conditional exposure odds model
on the left-hand side of equation (\ref{MIodds}) can be constructed
using logistic regression analysis with any functions of $\mathbf{y}_i$
and $\mathbf{x}_{oi}$ as independent variables. Since $X_{e,i} \perp S_i
\vert (\mathbf{Y}_i, \mathbf{X}_{oi})$ by design, then if the Gaussian linear
mixed model assumptions are satisfied, the induced conditional exposure
log-odds from equation (\ref{MIodds}) can be written
\begin{eqnarray}
&&-\frac{1}{2} \bigl\{ (\mathbf{Y}_i - \bolds{
\mu}_{1,i})^t \mathbf{V}_{1i}^{-1} (
\mathbf{Y}_i - \bolds{\mu}_{1,i}) - (\mathbf{Y}_i
- \bolds{\mu}_{0,i})^t \mathbf{V}_{0i}^{-1}
(\mathbf{Y}_i - \bolds{\mu}_{0,i}) \bigr\}
\nonumber
\\[-8pt]
\label{DirectMILogOdds}
\\[-8pt]
\nonumber
&&\qquad{}- \frac{1}{2} \log \biggl\{ \frac{\llvert  \mathbf
{V}_{1,i} \rrvert  }{ \llvert  \mathbf{V}_{0,i} \rrvert } \biggr\}
+ \log \biggl\{ \frac{\operatorname{pr}(X_{ei}=1 \vert \mathbf
{X}_{oi})}{\operatorname{pr}(X_{ei}=0 \vert \mathbf{X}_{oi})} \biggr\},
\nonumber
\end{eqnarray}
where $\bolds{\mu}_{x,i}=E(\mathbf{Y}_i \vert  X_{ei}=x,
\mathbf{x}_{oi})$, $\mathbf{V}_{x,i} = \operatorname{Var}( \mathbf{Y}_i
\vert  X_{ei}=x,
\mathbf{x}_{oi})=\mathbf{Z}_i \mathbf{D}_{xi} \mathbf{Z}_i + \sigma
_e^2 \mathbf{I}$.
If we assume homoscedasticity, then $\mathbf{V}_{1,i} = \mathbf
{V}_{0,i} =
\mathbf{V}_{i}$, and equation (\ref{DirectMILogOdds}) simplifies to
\begin{eqnarray}
&& \underbrace{\mathbf{Y}_i^t \mathbf{V}_{i}^{-1}
( \bolds{\mu }_{1, i} - \bolds{\mu}_{0, i} )}_{\mathrm{(a)}} \,-\,
\frac{1}{2} \underbrace {\bigl(\bolds{\mu}_{1,i}^t
\mathbf{V}_{i}^{-1} \bolds{\mu}_{1,i} - \bolds{
\mu}_{0,i}^t \mathbf{V}_{i}^{-1} \bolds{
\mu}_{0,i}\bigr)}_{\mathrm{(b)}}
\nonumber
\\[-8pt]
\label{DirectMILogOddsLDA}\\[-8pt]
\nonumber
&& \qquad{}+\underbrace{\log \biggl\{ \frac{\operatorname{pr}(X_{ei}=1 \vert \mathbf{X}_{oi})}{\operatorname{pr}(X_{ei}=0
\vert \mathbf{X}_{oi})} \biggr
\}}_{\mathrm{(c)}}.\nonumber
\end{eqnarray}
In the Supplement A [\citet{supp}], we detail further simplifications with
balanced and complete data that we examine in Section~\ref{Simulation}
and that are motivated by the CAMP analysis whose design was nearly
balanced and complete. Briefly, with balanced and complete data, if
$\nu_{i,jk}$ is the $(j,k)${th} element of $\mathbf{V}_i^{-1}$ and
$\omega_{ij}(t_{ij})$ is the $x_{ei}$ effect\vspace*{1pt} at time $t_{ij}$ (i.e.,
$\mu_{1,ij} - \mu_{0,ij}$), (a) equals $\sum_{j=1}^{n} \sum
_{k=1}^{n} \nu_{i,jk} \cdot y_{ij} \cdot\omega_{ik}(t_{ik})$, (b)~equals\vspace*{1pt} $\sum_{j=1}^{n} \sum_{k=1}^{n} \nu_{i,jk} \cdot\mu
_{0,ij} \cdot\omega_{ik}(t_{ik}) + \sum_{j=1}^{n} \sum
_{k=1}^{n} \nu_{i,jk} \cdot\omega_{ij}(t_{ij}) \cdot\omega
_{ik}(t_{ik})$, and (c)~contains terms involving $\mathbf{x}_{oi}$
useful for predicting $X_{ei}$. Our approach to imputation is then to
directly model $[X_{ei} \vert \mathbf{X}_{oi}, \mathbf{Y}_i]$ with logistic
regression and to follow standard multiple imputation methods. We note
that the first two terms in equations (\ref{DirectMILogOdds}) and
(\ref{DirectMILogOddsLDA}), respectively, result in the functional
form of quadratic and linear discriminant analysis [\citet{Fisher1936}] that
are used in many classification analyses.

%

\section{Finite sampling operating characteristics} \label{Simulation}

The key motivator in outcome dependent sampling schemes is to obtain
nearly efficient inference at considerable cost savings by drawing and
analyzing small to modest sample sizes. Indeed, the CAMP study could
have realized considerable savings if it had only analyzed 250
genotypes, versus more than 500. As such, it is critical in application
of these design strategies to quantify the degree to which theoretical
results are realized in finite sample settings. \citet{pmid23409789} conducted such simulations, that are briefly summarized in the
\hyperref[sec1]{Introduction}. We now examine the CD$+$MI and D-MI estimation procedures
proposed in Section~\ref{Model} to explore: (1) the scenarios under
which MI does and does not improve estimation efficiency over a CD
analysis; and (2) the extent to which the study design continues to
improve efficiency if MI is the intended analytical strategy. 

\subsection{Population model}

We conducted simulation studies under several study designs and
population features motivated by the CAMP study and by studies with
similarly-balanced longitudinal follow-up. Results presented here
summarize 1000 replications per scenario. In each scenario, we
generated a cohort of $N$ subjects based on the model
\begin{eqnarray*}
Y_{ij} &= & \beta_0 + \beta_t
t_{ij} + \beta_g g_i + \beta_{gt}
g_i t_{ij} + \beta_c c_i +
b_{0i} + b_{1i} t_{ij} + \varepsilon_{ij},
\end{eqnarray*}
with $i \in\{1,2, \ldots, N \}$ denoting subject, $j \in\{1, 2, \dots
, 10\}$ denoting observation within subject, $t_{ij}$ an equally
spaced, balanced time covariate ranging from $-2$ to 2, $C_i$ a binary,
time-invariant covariate with $\operatorname{pr}(C_i=1)=0.5$, $G_i$ an
expensive, binary ``group'' or ``genotype'' variable with $\operatorname{pr}(G_i=1 \vert  C_i=c) = 0.4 + \delta_c c$, $(b_{i0}, b_{i1})$ the
random intercept and slope, and $\varepsilon_{ij}$ the measurement error.
Across all scenarios, $(\beta_0, \beta_t, \beta_{gt})=(5, 1.0,
0.75)$, the mean of the random effects and error distributions were 0,
and the standard deviations of the random intercept, the random slope
and the measurement error were $\sigma_0=5$, $\sigma_1=1.25$ and
$\sigma_e=5$, respectively. Additionally, $\rho= \operatorname{corr}(b_{0i}, b_{1i})=-0.25$.

We examined the relative efficiency of the designs and estimation
procedures as a function of the following: the $G_i$ effect size,
$\beta_g \in\{-2.5, -4.0 \}$, the strength of the $G_i \sim C_i$
relationship, $\delta_c \in\{0.15, 0.35, 0.55 \}$, the sample size
of the original cohort $N \in\{750, 2250\}$, and the impact of $C_i$
being a proxy for $G_i$ as opposed to being a confounder for the $G_i
\sim\mathbf{Y}_i$ relationship. In the last scenario, $\beta_c=0$
and $C_i$
is used to impute $G_i$ but is not included in the primary analysis
model. In all other scenarios, $\beta_c=1$ and $C_i$ is included as an
independent variable. Specifically, we examine five distinct scenarios
uniquely identified by $(N, \beta_g, \delta_c, \beta_c)$. Scenarios
studied are given by the following: (a)~$(750, -2.5, 0.15, 1.0)$, (b)~$(750, -4.0, 0.15, 1.0)$, (c) $(750, -2.5, 0.35, 1.0)$, (d) $(2250,
-2.5, 0.15, 1.0)$, and (e) $(750, -2.5, 0.55, 0.0)$.

\subsection{Study designs}
The substudies we sought to examine were those that sampled, on
average, 250 subjects for whom $G_i$ should be ascertained, again
motivated by the CAMP framework. For the random sampling (RS) design,
we took a simple random sample of 250 subjects at each replication. For
ODS designs based on the intercept (ods.i), slope (ods.s), and
bivariate intercept and slope (ods.b), we calculated subject-specific
intercepts and slopes based on the $N$ separate OLS regressions of the
response $Y_{ij}$ on time $t_{ij}$, and sampled subject $i$ with
probability that depended upon the region in which $\mathbf{Q}_i$ was
located. For ods.i and ods.s we split the distribution of the sampling
variable $\mathbf{Q}_i$ into three regions defined by the 12th and
88th percentiles of the population distribution. We then sampled
individuals with probability $\pi(\mathbf{q}_i)=\operatorname{pr}(S_i=1
\vert
\mathbf{Q}_i=\mathbf{q}_i)$ so that, on average, 90 subjects from
each of the
two outlying regions and 70 subjects from the central region were
included in the outcome dependent sample. Similarly for the ods.b
design, we sampled with probability so that 70 subjects were included
from the central rectangular region that contained 76 percent of the
population and 180 subjects were included from the outlying region
containing 24 percent of the population. See \citet{pmid23409789} for a description and a figure describing these sampling schemes.

\subsection{Analyses}
After subsampling from the original cohort of $N$, we conducted the CD
analysis by fitting the model with maximum ascertainment corrected
likelihood under the ODS designs or with standard maximum likelihood
(ML) under the RS design. To conduct multiple imputation analyses, we
estimated the multiple imputation model for $G_i$ in unsampled subjects
$\operatorname{pr}(G_i \vert \mathbf{y}_i, c_i, S_i=0)$ via approaches discussed
in Sections~\ref{MIApproachesCDMI} and \ref{MIApproachesDMI}.
Specifically, the imputation model for CD$+$MI analyses was estimated by
combining the CD analysis and the offsetted logistic regression
analysis of $g_i$ on $c_i$ in sampled subjects. The imputation model
for the D-MI approach was estimated with a regression model of $G_i$ on
predictors $\sum_{j} y_{ij}$, $\sum_{j} y_{ij} \cdot t_{ij}$ and
$C_i$ in sampled subjects. See the online supplementary materials [\citet{supp}] for
an explanation of why these independent variables were used in the
imputation model. The number $M$ of imputation samples used was based
on examination of the degrees of freedom that were calculated as
described in Section~\ref{MISection} and with the intention of
conducting sufficient imputation analyses so that the $t$-statistics
associated with all parameter estimates were well approximated by
normal distributions for all parameters. When $N=750$, $M=25$ was used;
when $N=2250$,~$M=35$.

\subsection{Results}

Because the models were properly specified, all estimation procedures
were observed to be approximately valid with observed biases in
parameter estimates less than $5\%$ and observed biases in standard
errors less than $10\%$ (not shown).
\begin{sidewaystable}
\tablewidth=\textwidth
\caption{Relative efficiency: Results show ratios of the empirical
variance of the RS design with standard CD analyses to the empirical
variance of all other study design and analysis procedures across 1000
replicates. The designs ods.i, ods.s and ods.b are ODS designs with
sampling based on the intercept, slope, and both intercept and slope of
subject-specific ordinary least squares regression of $Y_{ij}$ on
$t_{ij}$. For each parameter we show columns that correspond to CD,
CD$+$MI and D-MI analyses, respectively. In scenario (\textup{e}) we do not
estimate $\beta_c$, as $C_i$ is not included in the final model but is
only used for $G_i$~imputation} \label{RelEffTab}
\begin{tabular*}{\tablewidth}{@{\extracolsep{\fill}}lccccccc@{}}
\hline
& $\bolds{N}$, $\bolds{\beta_g}$, $\bolds{\delta_c}$, $\bolds{\beta_c}$ & \textbf{Design} & $\bolds{\beta_0}$ &
$\bolds{\beta_t}$ & $\bolds{\beta_g}$ & $\bolds{\beta_{gt}}$ & $\bolds{\beta_c}$ \\
\hline
(a) &750, $-$2.5, 0.15, 1.0 & RS & 1.00, 1.88, 1.90 & 1.00, 1.68, 1.64 &
1.00, 1.02, 1.03 & 1.00, 1.13, 1.09 & 1.00, 2.66, 2.65 \\
&& ods.i & 2.18, 2.63, 2.63 & 0.89, 1.37, 1.35 & 2.11, 2.20, 2.19 &
0.88, 0.94, 0.92 & 1.99, 2.64, 2.64 \\
&& ods.s & 1.02, 1.89, 1.90 & 2.01, 2.32, 2.27 & 1.00, 1.00, 1.02 &
1.87, 2.01, 1.96 & 1.03, 2.61, 2.62 \\
&& ods.b & 1.82, 2.42, 2.41 & 1.64, 1.97, 1.97 & 1.75, 1.79, 1.82 &
1.52, 1.59, 1.59 & 1.72, 2.67, 2.65 \\[3pt]
(b) &750, $-$4.0, 0.15, 1.0 & RS & 1.00, 1.90, 1.92 & 1.00, 1.65, 1.67 &
1.00, 1.20, 1.21 & 1.00, 1.14, 1.16 & 1.00, 2.65, 2.59 \\
&& ods.i & 1.79, 2.17, 2.14 & 1.02, 1.61, 1.57 & 1.65, 1.99, 1.96 &
1.01, 1.07, 1.04 & 1.83, 2.27, 2.20 \\
&& ods.s & 1.01, 1.85, 1.83 & 2.30, 2.71, 2.74 & 0.91, 1.06, 1.05 &
2.19, 2.35, 2.36 & 1.00, 2.49, 2.48 \\
&& ods.b & 1.57, 2.13, 2.10 & 2.03, 2.44, 2.42 & 1.43, 1.57, 1.57 &
1.85, 1.98, 1.93 & 1.79, 2.53, 2.46 \\[3pt]
(c) &750, $-$2.5, 0.35, 1.0 & RS & 1.00, 1.90, 1.91 & 1.00, 1.61, 1.51 &
1.00, 1.05, 1.04 & 1.00, 1.23, 1.15 & 1.00, 2.26, 2.27 \\
&& ods.i & 2.03, 2.62, 2.62 & 1.00, 1.56, 1.48 & 1.95, 2.09, 2.12 &
0.96, 1.15, 1.08 & 1.90, 2.37, 2.39 \\
&& ods.s & 1.13, 2.06, 2.05 & 2.10, 2.53, 2.53 & 1.01, 1.06, 1.07 &
2.07, 2.33, 2.33 & 1.00, 2.28, 2.28 \\
&& ods.b & 1.89, 2.51, 2.51 & 1.88, 2.28, 2.24 & 1.71, 1.81, 1.78 &
1.84, 2.02, 1.95 & 1.67, 2.45, 2.41 \\[3pt]
(d)& 2250, $-$2.5, 0.15, 1.0 & RS & 1.00, 2.97, 3.01 & 1.00, 2.03, 2.00 &
1.00, 1.07, 1.07 & 1.00, 1.14, 1.11 & 1.00, 5.83, 5.79 \\
&& ods.i & 2.06, 4.69, 4.67 & 0.99, 1.97, 1.89 & 1.76, 2.01, 2.01 &
0.95, 1.11, 1.07 & 1.89, 5.75, 5.74 \\
&& ods.s & 0.98, 2.85, 2.89 & 2.12, 3.75, 3.70 & 0.92, 0.95, 0.97 &
2.05, 2.44, 2.39 & 0.86, 5.61, 5.52 \\
&& ods.b & 1.65, 3.98, 4.07 & 1.83, 3.25, 3.21 & 1.52, 1.57, 1.60 &
1.81, 2.02, 1.98 & 1.53, 5.76, 5.50 \\[3pt]
(e)& 750, $-$2.5, 0.55, 0.0 & RS & 1.00, 1.71, 1.60 & 1.00, 1.79, 1.59 &
1.00, 1.50, 1.37 & 1.00, 1.52, 1.32 & \\
&& ods.i & 1.95, 2.33, 2.33 & 1.03, 1.64, 1.49 & 1.98, 2.29, 2.29 &
0.92, 1.39, 1.20 & \\
&& ods.s & 1.04, 1.65, 1.58 & 1.99, 2.36, 2.33 & 0.99, 1.46, 1.36 &
2.03, 2.37, 2.33 & \\
&& ods.b & 1.77, 2.16, 2.09 & 1.80, 2.21, 2.17 & 1.75, 2.06, 1.96 &
1.77, 2.17, 2.08 & \\
\hline
\end{tabular*}
\end{sidewaystable}

Table~\ref{RelEffTab} shows the efficiency of each design and analysis
procedure combination relative to the RS design and standard CD maximum
likelihood analysis. Relative efficiency is defined as the empirical
variance under RS plus CD analyses divided by the empirical variance
under each other design and estimation procedure. Note that the CD$+$MI
and D-MI approaches perform similarly for nearly all
parameter-by-scenario combinations. In scenario (a) we observe that for
$\beta_g$ and $\beta_{gt}$ the impact of the study design far
outweighs the impact of multiply imputing $G_i$. For example, using CD
analyses to estimate $\beta_{gt}$, the ods.s design improves
estimation efficiency by 87 percent over RS, but adding multiple
imputation to the CD analysis by using the CD$+$MI approach improves
efficiency only by an additional $7.4$ percent ($2.01/1.87=1.074$).
However, if interest is in estimates of $\beta_c$, which correspond to
$C_i$, a covariate that is available in everyone, the impact of
multiple imputation outweighs the study design. Notice that with both
the CD$+$MI and D-MI approaches all designs have a relative efficiency
for $\beta_c$ of approximately 2.6--2.7 compared to random sampling
with CD analyses. For estimates of $\beta_0$ and $\beta_t$, the study
design and multiple imputation-based analyses independently contributed
to optimal estimation efficiency.
\begin{figure}[b]\vspace*{-5pt}

\includegraphics{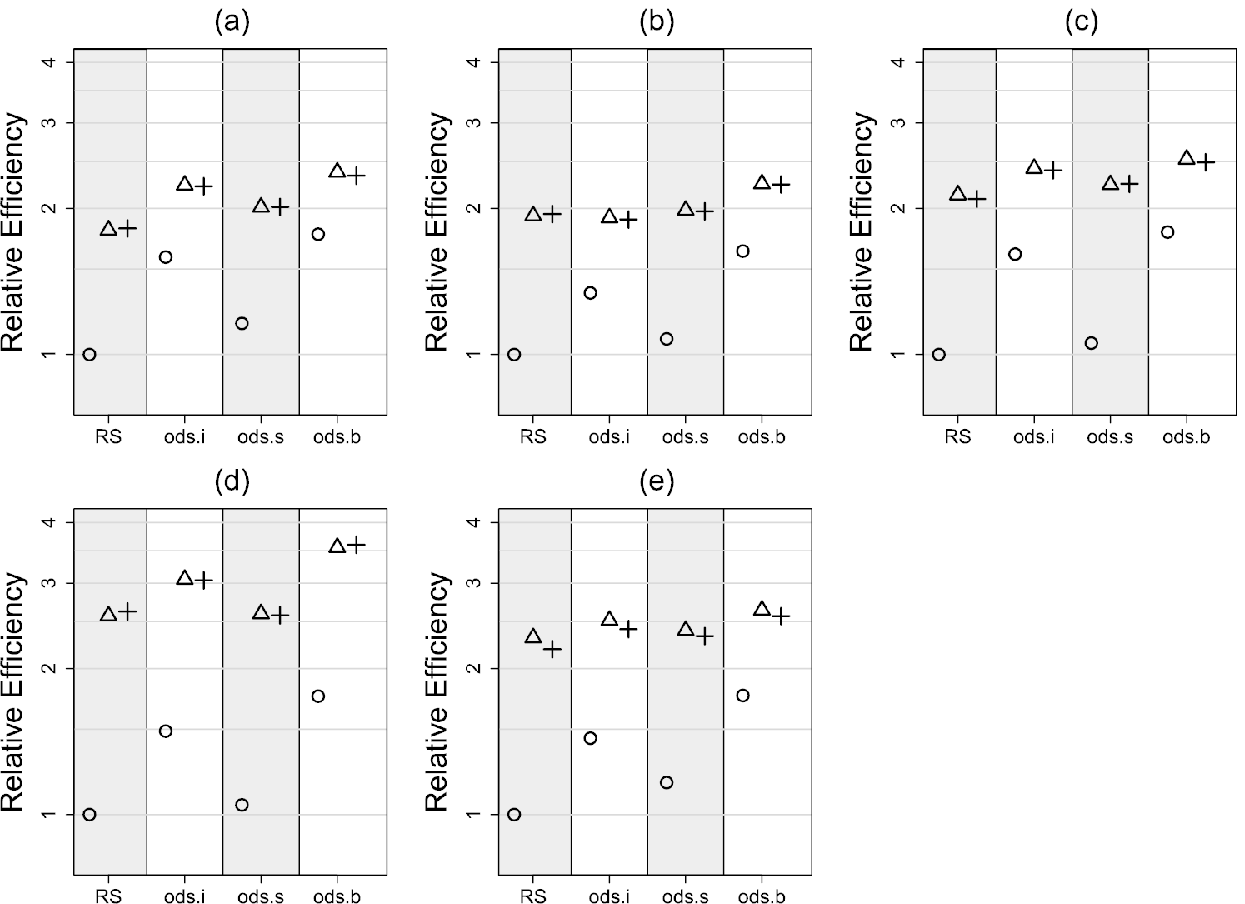}
\vspace*{-5pt}
\caption{Relative efficiency for estimating the predicted value at the
end of the study period $\mu_{i,10}= E(Y_{i,10} \vert  G_i=1, C_i=1,
t_{ij}=2)$ for all design and analysis procedure combinations versus RS
and standard CD analyses based on 1000 replications. Symbol $\mathrm{o}$ denotes
CD analyses, $\triangle$~denotes CD$+$MI analyses, and $+$ denotes D-MI
analyses. Parameter values \textup{(a)}--\textup{(e)} are given in Table~\protect\ref{RelEffTab}.}
\label{RelEffPlot}
\end{figure}

Scenarios (b), (c), (d) and (e) provide some insight into how the
results shown in scenario (a) depend upon population data features. We
used these scenarios specifically to examine the extent to which MI
adds to the optimal study design with CD analyses and we now focus our
discussion exclusively on $\beta_g$ and $\beta_{gt}$. Comparing
results from scenario (b) to (a), we observed that the impact of MI is
somewhat greater when the $G_i$ effect size is larger. Whereas in
scenario (b), when estimating $\beta_g$, CD$+$MI was 20 percent more
efficient than CD for the optimal ods.i design $(1.99/1.65=1.20)$, in
scenario (a) it was only 4 percent more efficient $(2.20/2.11=1.04)$.
As shown by comparing results from scenarios (c) and (d) to~(a), we
observe that MI appears to add modest additional precision to the
optimal design when the $G_i \sim C_i$ relationship is stronger and
when the original cohort size is larger. Finally, in scenario (e) we
observed that when $C_i$ is a proxy for $G_i$ rather than a confounder
and when the $G_i \sim C_i$ relationship is relatively strong with
$\delta_c=0.55$, adding MI to the optimal design led to larger
efficiency gains for $\beta_g$ and $\beta_{gt}$. For example, the
relative efficiency of CD$+$MI relative to CD analyses for the optimal
designs for $\beta_g$ and $\beta_{gt}$ were $2.29/1.98=1.16$ and
$2.37/2.03=1.17$, respectively.

Multiple imputation resulted in substantial efficiency improvements
over CD analysis for estimates of $(\beta_0, \beta_t, \beta_c)$, but
had a far smaller impact on estimation efficiency for $(\beta_g, \beta
_{gt})$. Figure~\ref{RelEffPlot} shows the relative efficiency for
estimating the mean value at the end of the study period for those with
$(G_i,C_i)=(1,1)$, $\widehat{\mu}_{i,10}=E(Y_{ij} \vert  G_i=1, C_i=1,
t_{ij}=2)$ under all scenarios. By combining parameter estimates to
obtain the linear predictor estimate we observed that in all scenarios
and for all study designs, CD$+$MI and D-MI analyses are substantially
more efficient than CD analyses. That is, MI improved estimation
efficiency dramatically, and the study design itself had a more modest
impact. However we also note that the ods.b design is the most
efficient design in all scenarios for estimating the end-of-study mean
value. Even though ods.b was not the optimal design for any single
parameter (see Table~\ref{RelEffTab}), it is reasonably efficient for
all parameters, which is beneficial if more than one parameter is of
interest. In contrast, the ods.s and ods.i designs were efficient for
individual parameters but were inefficient for other parameters.\vspace*{-3pt}

\section{CAMP data analysis} \label{CAMP}

In this section we analyze the CAMP data using different subsampling
designs both with and without imputation. Our goal is to empirically
compare the research efficiency of candidate designs, and we have the
complete data against which we can benchmark performance. Since our
simulation study showed that CD$+$MI and D-MI approaches are similar, we
focus our presentation on only one imputation approach (CD$+$MI). A total
of 555 subjects had sufficient covariate and genotype data available,
and we operate under the assumption that stored blood samples are
available for all participants, although study resources only permit
genotyping 250. Thus, approximately 250 genotypes are used at each of
30 replications of each study design. We report results based on the
average estimates and (co)variances. Similar to the simulations, we
consider four designs: random subsampling of 250 children (RS) and
three ODS designs. To create the ODS designs, we first compute all
estimated intercepts and slopes from subject-specific simple linear
regressions of post-bronchodilator percent predicted FVC (FVC\%) on
time since randomization. Sampling was then based on the following: the
estimated intercept (ods.i), the estimated slope (ods.s), or the
estimated intercept and slope jointly (ods.b). In order to obtain 250
subjects, the cutoff points that define strata in the ods.i and ods.s
designs are given by the 16th and 84th percentiles of the
original cohort. We sampled with probability 1 subjects at or below the
16th percentile and at or above the 84th percentile, and with
probability 0.19 all subjects falling in the central $68$\% region. For
ods.b, we sampled with probability 0.19 all subjects who fell in the
central $68$\% region of the joint intercept and slope distribution in
the original cohort and with probability 1 all of those falling outside
this region. Table $\ref{CAMPDemog}$ shows the characteristics of the
CAMP cohort from which we subsampled for the ODS studies.

\begin{table}
\tablewidth=260pt
\caption{Demographic and other characteristics of children
participating in the CAMP with genotype and covariate data available.
Continuous variables are summarized with the 10th, 50th and
90th percentiles, and categorical variables other than site are
summarized with proportions} \label{CAMPDemog}
\begin{tabular*}{260pt}{@{\extracolsep{\fill}}lc@{}}
 \hline
\textbf{Variable} & \\
\hline
Cohort size ($N$) & 555 \\
\quad  Albuquerque & 41 \\
\quad  Baltimore & 71 \\
\quad  Boston & 72 \\
\quad  Denver & 64 \\
\quad  San Diego & 68 \\
\quad  Seattle & 80 \\
\quad  Saint Louis & 91 \\
\quad  Toronto & 68 \\
Age at randomization (years) & 6.23, 8.81, 11.71 \\
Male gender & 0.65 \\
Black race & 0.10 \\
Other (noncaucasian) race & 0.26 \\
Randomized treatment & \\
\quad  Placebo & 0.50 \\
\quad  Budesonide & 0.32 \\
\quad  Nedocromil & 0.17 \\
IL10 variant allele & 0.50 \\
Observations per subject & 9, 10, 10 \\
Follow-up time (years) & 3.85, 3.99, 4.1 \\
Post-bronchodilator percent predicted & 92, 105, 116 \\
\hline
\end{tabular*}
\end{table}
\begin{sidewaystable}
\tablewidth=\textwidth
\tabcolsep=0pt
\caption{CAMP results: estimated summaries and standard error
estimates (in parentheses) based on 30 replications of each study
design. At each replication, twenty imputation samples were used for
the CD$+$MI analyses. We do not include the standard errors for variance
components with the CD$+$MI approach because the lme4 package [\citet{BaMa2010}] does not provide them. Although site effects are not
shown, they were included as fixed effects in regression analyses. The
estimated mean row corresponds to the estimated, end-of-study mean
value for the population of white, 12 year old girls, with VAs who were
randomized to placebo treatment and who lived in Baltimore. The
original cohort column displays results from the analysis of the full
cohort of 555 participants}\label{CAMPResults}
\begin{tabular*}{\tablewidth}{@{\extracolsep{\fill}}ld{3.8}d{3.8}d{3.8}d{3.8}d{3.8}d{3.8}d{3.8}d{3.8}d{3.8}@{\hspace*{-1pt}}}
 \hline
\textbf{Variable} & \multicolumn{1}{c}{\multirow{2}{33pt}{\centering\textbf{Original} \textbf{cohort}}}& \multicolumn{2}{c}{\textbf{RS}} &\multicolumn
{2}{c}{\textbf{ods.s}} &\multicolumn{2}{c}{\textbf{ods.i}} &\multicolumn{2}{c@{}}{\textbf{ods.b}} \\[-4pt]
& & \multicolumn{2}{l}{\hrulefill} & \multicolumn{2}{l}{\hrulefill}& \multicolumn{2}{l}{\hrulefill}& \multicolumn{2}{l@{}}{\hrulefill}\\
&  & \multicolumn{1}{c}{\textbf{CD}} & \multicolumn{1}{c}{\textbf{CD$\bolds{+}$MI}} &\multicolumn{1}{c}{\textbf{CD}} &
\multicolumn{1}{c}{\textbf{CD$\bolds{+}$MI}} &\multicolumn{1}{c}{\textbf{CD}} & \multicolumn{1}{c}{\textbf{CD$\bolds{+}$MI}} &
\multicolumn{1}{c}{\textbf{CD}} & \multicolumn{1}{c}{\textbf{CD$\bolds{+}$MI}} \\
\hline
\multicolumn{10}{@{}c@{}}{Primary summaries} \\
\multicolumn{10}{l}{Budesomide (vs placebo) at all times} \\
\quad No VAs & -2.11 \ (1.16) & -1.57  \ (1.73) & -2.09\  (1.46)
& -3.65\  (1.73) & -2.92\  (1.45) & -2.39\  (1.41) & -2.73\  (1.29) & -2.65
\ (1.56) & -2.68\  (1.34)\\
\quad With VAs & 3.29\  (1.24) & 3.08\  (1.86) & 3.08\  (1.52) &
4.12\  (1.92) & 4.18\  (1.54) & 3.99\  (1.55) & 3.95\  (1.38) & 3.51\  (1.67) &
3.69\  (1.38)\\
\quad Difference & 5.40\  (1.70) & 4.65\  (2.54) & 5.17\  (2.43)
& 7.78\  (2.57) & 7.10\  (2.42) & 6.39\  (2.10) & 6.68\  (2.03) & 6.16\  (2.34) &
6.37\  (2.10)\\[3pt]
\multicolumn{10}{@{}l@{}}{Nedocrimil  (vs placebo) at all times} \\
\quad  No VAs & -0.77\  (1.17) & -0.62\  (1.73) & -0.56\  (1.46)
& -2.96\  (1.73) & -1.59\  (1.46) & -1.11\  (1.41) & -0.56\  (1.24) & -2.16 \
(1.51) & -0.96\  (1.31)\\
\quad  With VAs & 0.69\  (1.10) & 0.73\  (1.63) & 0.54\  (1.39) &
0.42\  (1.63) & 1.45\  (1.36) & 0.31\  (1.31) & 0.64\  (1.21) & -0.02\  (1.36) &
0.92\  (1.20)\\
\quad  Difference & 1.46\  (1.61) & 1.35\  (2.39) & 1.10\  (2.36)
& 3.38\  (2.41) & 3.04\  (2.33) & 1.42\  (1.94) & 1.20\  (1.87) & 2.14\  (2.08) &
1.88\  (1.95)\\[3pt]
\multicolumn{10}{@{}l@{}}{Time trend  (per year) irrespective of treatment} \\
\quad  No VAs & 0.14\  (0.16) & 0.11\  (0.23) & 0.09\  (0.19) &
0.14\  (0.17) & 0.10\  (0.16) & -0.04\  (0.22) & 0.09\  (0.18) & 0.19\  (0.18) &
0.13\  (0.17)\\
\quad  With VAs & -0.25\  (0.15) & -0.19\  (0.23) & -0.19
\ (0.19) & -0.19\  (0.17) & -0.21\  (0.16) & -0.38\  (0.22) & -0.21\  (0.18) &
-0.25\  (0.18) & -0.24\  (0.16) \\
\quad  Difference & -0.39\  (0.22) & -0.30\  (0.33) & -0.27
\ (0.31) & -0.33\  (0.24) & -0.31\  (0.24) & -0.35\  (0.31) & -0.30\  (0.29) &
-0.44\  (0.26) & -0.37\  (0.25)\\[3pt]
\multicolumn{10}{@{}l@{}}{IL10 (VA vs no VA) in the placebo arm at baseline
and year 4} \\
\quad
$t_{ij=0}$ & -1.65\  (1.15) & -1.67\  (1.70) & -1.78 \
(1.69) & -2.20\  (1.65) & -2.13\  (1.67) & -1.72\  (1.30) & -1.90\  (1.32) &
-1.29\  (1.46) & -1.50\  (1.39)\\
\quad $t_{ij=4}$ & -3.2\  (1.17) & -2.87\  (1.73) & -2.88 \
(1.71) & -3.51\  (1.68) & -3.35\  (1.68) & -3.11\  (1.5) & -3.12\  (1.52) &
-3.05\  (1.52) & -2.98\  (1.48)\\[3pt]
Estimated mean & 106.33\  (1.51) & 106.86\  (2.32) & 106.47\  (1.64) & 107.95  \
(2.20) & 106.23\  (1.62) & 106.69\  (1.99) & 106.42\  (1.63) & 106.79\  (1.97)
& 106.47\  (1.59) \\
\hline
\end{tabular*}
\end{sidewaystable}

\setcounter{table}{2}
\begin{sidewaystable}
\tablewidth=\textwidth
\tabcolsep=0pt
\caption{(Continued)}
\begin{tabular*}{\tablewidth}{@{\extracolsep{\fill}}ld{3.8}d{3.8}d{3.8}d{3.8}d{3.8}d{3.8}d{3.8}d{3.8}d{3.8}@{\hspace*{-1pt}}}
 \hline
\textbf{Variable} & \multicolumn{1}{c}{\multirow{2}{33pt}{\centering\textbf{Original} \textbf{cohort}}}& \multicolumn{2}{c}{\textbf{RS}} &\multicolumn
{2}{c}{\textbf{ods.s}} &\multicolumn{2}{c}{\textbf{ods.i}} &\multicolumn{2}{c@{}}{\textbf{ods.b}} \\[-4pt]
& & \multicolumn{2}{l}{\hrulefill} & \multicolumn{2}{l}{\hrulefill}& \multicolumn{2}{l}{\hrulefill}& \multicolumn{2}{l@{}}{\hrulefill}\\
&  & \multicolumn{1}{c}{\textbf{CD}} & \multicolumn{1}{c}{\textbf{CD$\bolds{+}$MI}} &\multicolumn{1}{c}{\textbf{CD}} &
\multicolumn{1}{c}{\textbf{CD$\bolds{+}$MI}} &\multicolumn{1}{c}{\textbf{CD}} & \multicolumn{1}{c}{\textbf{CD$\bolds{+}$MI}} &
\multicolumn{1}{c}{\textbf{CD}} & \multicolumn{1}{c}{\textbf{CD$\bolds{+}$MI}} \\
\hline
\multicolumn{10}{@{}c@{}}{Other mean model parameters} \\
Male  (vs female) & -1.14\  (0.72) & -1.47\  (1.08) & -1.22\  (0.73) & -1.47 \
(1.07) & -1.13\  (0.72) & -0.71\  (0.86) & -1.16\  (0.72) & -1.19\  (0.90) &
-1.21\  (0.72)\\
Black  (vs white) & 0.51\  (1.21) & 0.52\  (1.87) & 0.53\  (1.25) & 1.22 \
(1.85) & 0.76\  (1.23) & 1.19\  (1.56) & 0.47\  (1.24) & 1.88\  (1.50) & 0.67 \
(1.23)\\
Other  (vs white) & -0.81\  (0.98) & -0.95\  (1.44) & -0.74\  (0.99) & -1.31 \
(1.44) & -0.59\  (1.00) & -0.01\  (1.15) & -0.71\  (0.99) & -0.32\  (1.20) &
-0.61\  (0.99)\\
Age  ($t_{ij}=0$) & -0.21\  (0.17) & -0.23\  (0.26) & -0.22\  (0.17) & -0.40 \
(0.26) & -0.23\  (0.17) & -0.50\  (0.21) & -0.22\  (0.17) & -0.39\  (0.22) &
-0.22\  (0.17)\\[3pt]
\multicolumn{10}{@{}c@{}}{Variance components} \\
$\log(\sigma_0)$ & 2.19 & 2.18\  (0.05) & 2.19 & 2.16\  (0.05) & 2.18 &
2.18\  (0.04) & 2.18 & 2.18\  (0.04) & 2.18 \\
$\log(\sigma_1)$ & 0.84 & 0.85 \ (0.06) & 0.84 & 0.84\  (0.05) & 0.84 &
0.83\  (0.05) & 0.84 & 0.84\  (0.05) & 0.84\\
$\frac{\log(1+\rho)}{\log(1-\rho)}$ & -1.70 & -1.13\  (0.15) & -1.70 &
-1.06\  (0.12) & -1.70 & -1.11\  (0.12) & -1.70 & -1.08\  (0.12) & -1.69 \\
$\log(\sigma_e)$ & 1.55 & 1.54\  (0.02) & 1.55 & 1.60\  (0.02) & 1.55 &
1.60\  (0.02) & 1.55 & 1.62\  (0.02) & 1.55 \\
\hline
\end{tabular*}
\end{sidewaystable}

The primary scientific goals of the CAMP analysis are to examine the
treatment effects within subgroups defined by the presence or absence
of a variant allele (VA) on the fourth locus of the IL10 gene, and to
examine the difference in lung growth between those with and without a
VA. Three-way interactions (IL10 $\times$ medication~$\times$
$t_{ij}$) were explored, however, we only report results from two-way
interactions. Thus, the fitted model for this analysis was
\begin{eqnarray*}
E[ y_{ij} \vert  X_i ] & = & \beta_0 +
\beta_1 t_{ij} + \beta_2 \cdot
\mathit{bud}_i + \beta_3 \cdot \mathit{ned}_i +
\beta_4 \cdot \mathrm{IL}10_i + \beta_5 \cdot
\mathit{bud}_i \cdot \mathrm{IL}10_i
\\
& &{}+ \beta_6 \cdot \mathit{ned}_i \cdot
\mathrm{IL}10_i + \beta_7 \cdot t_{ij} \cdot
\mathrm{IL}10_i + \beta_C \cdot \mathit{covariates}_{ij}.
\end{eqnarray*}
The covariates that represent the key biomedical questions include the
following: the binary time invariant IL10 SNP ($\mathit{snp}_i$); time since
randomization ($\mathbf{t}_i = \{t_{i1}, \dots, t_{in_i} \}$); Budesonide
($\mathit{bud}_i$) and Nedocromil ($\mathit{ned}_i$) treatments (with placebo being the
reference); and pairwise interactions between IL10 and the other
variables. As described in Section~\ref{MISection}, the imputation
approaches required a model for the predictor of interest,
$X_{ei}=\mathit{snp}_i$, in order to impute its value for subjects not selected
for the subsample ($S_i=0$). Therefore, the CD$+$MI analysis procedure
required estimation of a marginal exposure distribution (i.e., $[X_{ei}
\vert \mathbf{X}_{oi}, S_i=1]$), and in that model, $\mathit{race}_i$, $\mathit{site}_i$,
$\mathit{gender}_i$, $\mathit{bud}_i$ and $\mathit{ned}_i$ were used as independent variables
($\mathbf{X}_{oi}$) in an additive logistic regression model.

Table~\ref{CAMPResults} shows CAMP regression summaries based on the
original cohort analysis using all subjects ($N=555$), and on eight
combinations of subsampling designs with and without imputation, where
only $N \approx 250$ children were included in a subsample. We provide
the key summaries that specifically address the primary research
questions, but interested readers may look to online supplementary
materials [\citet{supp}] for all longitudinal model regression estimates and
interactions used to generate the summaries. Specifically, we focus on
medication effects and time trends within subgroups defined by presence
or absence of an IL10 variant, the difference in expected FVC between
those with and without an IL10 variant at baseline ($t_{ij}=0$) and at
the end of the study ($t_{ij}=4$) for subjects on placebo treatment,
and the end-of-study predicted mean value.

In the original cohort analysis we observed the following associations
that were statistically significant at the $\alpha=0.05$ level: (1)
for subjects with an IL10 variant, the expected FVC\% was estimated to
be 3.29 (1.24) units higher across all times in those randomized to
Budesomide compared to placebo; (2) the effect of Budesomide compared
to placebo was 5.40 (1.70) units higher in those with an IL10 variant
than in those without an IL10 variant; and (3) at the end of the study
($t_{ij}= 4$), those with an IL10 variant were estimated to have FVC\%
values that were 3.20 (1.17) units lower than those without an IL10
variant. Our interest is in the impact of subsampling design choices,
so a natural option to consider is a simple random sample. However,
although the random sampling design produced point estimates that were
similar to results from the original cohort, none of the full
cohort-based associations would be considered statistically significant
using the RS design. In contrast, all ODS designs detected the three
significant effects seen in the original cohort, demonstrating the
potential efficiency gains though use of biased sampling in a
resource-limited environment.

Furthermore, for all designs the use of imputation (CD$+$MI analysis)
improved estimation efficiency of key parameters. For example, when
sampling using ods.b, the standard error for the Budesomide versus
placebo contrast was 1.67 under the CD analysis, 1.38 under the CD$+$MI
analysis and 1.24 for the original cohort analysis. Such efficiency
gains due to MI were also observed in all coefficient estimates for the
other baseline covariates measured on all subjects (e.g., age, race and
gender).
In contrast, and consistent with simulations, CD${}+{}$MI did not produce appreciably smaller
estimates of uncertainty than CD analyses for parameters that capture (retrospectively
ascertained) IL10 effects and interactions.  For example, under the ods.b design,
the standard error estimate for the IL10 VA association with FVC\%
in the placebo arm at $t_{ij}=4$ was 1.52 and 1.48 with CD and CD${}+{}$MI analyses,
respectively.
Similarly, the standard error estimate for the difference in the time trends between those with and without
the IL10 VA was 0.26 and 0.25 with CD and CD${}+{}$MI analyses, respectively.

Finally, for many parameters, the combination of subsampling and the
use of imputation was able to recover a large fraction of the
information present in the original cohort but with less than half the
cost in terms of number of subjects for whom covariates would be
ascertained. For example, all estimators produced quite similar
estimates of the predicted mean value at the end of the study, ranging
from 106.33 to 107.95, and the ods.b plus CD$+$MI combination estimated
the standard error to be 1.59, only slightly higher than the 1.51
estimated from the original cohort. In summary, the CAMP analysis
illustrates that targeted subsampling is typically more efficient than
simple random sampling, and that using all available data is also
beneficial and can be easily accomplished through imputation of data
for those subjects not selected in a given subsample. We recommend that
future ancillary studies of existing longitudinal cohorts consider the
benefits of directed sampling coupled with efficient analysis.

\section{Discussion} \label{Discussion}

The CAMP longitudinal clinical trial was conducted in an era when
genotyping was more expensive than today. Owing to ongoing interest in
treatment heterogeneity, it is of interest to examine whether treatment
effectiveness varies across genotype. Because this would be a secondary
aim of most trials, it makes sense economically to conduct the trial,
obtain response trajectories and test for overall treatment
effectiveness first. Depending on what is learned through those primary
investigations, investigators---or their colleagues---may then wish to
move ahead with other exposure assessments to examine exposure effects
or treatment-by-exposure interactions. Such data could be used for
confirmatory analyses or, more likely, for pilot or preliminary data in
an exploratory model. In these kinds of settings, especially, cost
effectiveness is critical, and can make the difference between a study
being viable or not.

To address such problems, in this manuscript we discussed novel
statistical approaches to the combination both of ODS designs and of
efficient analyses for longitudinal continuous response data. We
observed that MI-based approaches can improve efficiency dramatically
over CD analyses for parameters corresponding to estimation targets
involving covariates that were not imputed (e.g., demographics and the
estimated mean value in CAMP). Efficiency improvements were more modest
for the coefficients of imputed covariates (e.g., the VA by time
interaction under the $ods.s$ design in CAMP), although such results
can be influenced by data features (e.g., effect size in simulation).
Importantly, we also observed that, even when MI is a default
analytical choice, ODS designs can still improve efficiency
dramatically in targets associated (directly or indirectly through
interactions) with the retrospectively ascertained covariate.

Because this manuscript discusses what we believe are new study
designs, we were not able to analyze data directly from such a study.
Such studies have yet to be conducted. Instead, to describe the
characteristics of the designs and estimators, we replicated simulated
substudies from CAMP. While this may not appear to be ideal at first,
it allowed us to explore alternative CAMP substudy designs and did not
lock us in to a single design. 

The two MI strategies, CD-MI and D-MI, approach parameter estimation in
somewhat different ways, even in the context of the overall MI
framework. Specifically, both approaches require careful consideration
of two model specifications. Whereas the outcome model $[\mathbf{Y}_i
\vert
X_{ei}, \mathbf{X}_{oi}]$ is common to both strategies, CD$+$MI requires the
direct specification of a marginal exposure model $[X_{ei} \vert \mathbf
{X}_{oi}]$ and D-MI requires the direct specification of the fully
conditional exposure model $[X_{ei} \vert \mathbf{X}_{oi}, \mathbf
{Y}_i]$. We
believe that each approach has an important advantage. In a relative
way, CD$+$MI may be considered advantageous because the marginal exposure
model is likely to be relatively simple as compared to the conditional
exposure model, and so the focus of analysis with CD$+$MI is on the
outcome model. The conditional exposure model that is directly
specified with D-MI is likely to involve additional consideration of
the functional form of a time-varying (response) variable toward
prediction of a time-fixed exposure variable. In contrast, the D-MI may
be considered more flexible because the outcome and imputation models
are decoupled. As compared to CD$+$MI, it could potentially be more
robust to misspecification of the outcome model.

Finally, a rigorous evaluation of competing approaches (e.g., inverse
probability weighting) is next in this line of research. A key reason
we have not pursued that here is that we are primarily interested in
situations wherein a full likelihood approach for both estimation and
inference is of interest. The IPW approaches step out of that paradigm,
instead relying on sandwich-type variance estimators, making the
comparison among the methods more complex. Other areas of future
research that specifically pertain to the imputation approaches involve
extensions of the exposure variable to continuous, ordinal and
time-varying data. We also intend to explore imbalanced time-varying
covariates, unequal cluster sizes, general patterns of missing data/dropout, mean model misspecification and imputation
model misspecification.

\section*{Acknowledgments}

The authors would like to thank the referees, the Associate
Editor and the Editor for their carefully considered feedback that
improved the manuscript appreciably. The work was conducted in part
using the resources of the Advanced Computing Center for Research and
Education at Vanderbilt University, Nashville, TN. The Childhood Asthma
Management Program trial and CAMP Continuation Study were supported by
contracts N01-HR-16044, 16045, 16046, 16047, 16048, 16049, 16050, 16051
and 16052 with the National Heart, Lung, and Blood Institute and
General Clinical Research Center grants M01RR00051, M01RR0099718-24,
M01RR02719-14 and RR00036 from the National Center for Research
Resources. The CAMP Genetics Ancillary Study is supported by grants
U01HL075419, U01HL65899 and P01HL083069 from the National Heart Lung
and Blood Institute.

\begin{supplement}[id=suppA]
\sname{Supplement A}
\stitle{D-MI Derivation for the model used in simulation}
\slink[doi]{10.1214/15-AOAS826SUPPA} 
\sdatatype{.pdf}
\sfilename{aoas826\_suppa.pdf}
\sdescription{Derivation of the D-MI imputation mo\-del used in
simulations (in Section~\ref{MIApproachesDMI}).}
\end{supplement}

\begin{supplement}[id=suppB]
\sname{Supplement B}
\stitle{CAMP Results: Parameter and uncertainty estimates}
\slink[doi]{10.1214/15-AOAS826SUPPB} 
\sdatatype{.pdf}
\sfilename{aoas826\_suppb.pdf}
\sdescription{Results from the CAMP analysis that were used to derive
the summaries in Table~\ref{CAMPResults}.}
\end{supplement}


\printaddresses

\begin{thebibliography}{40}


\bibitem[\protect\citeauthoryear{Bates and Maechler}{2010}]{BaMa2010}
\begin{bmisc}[author]
\bauthor{\bsnm{Bates},~\bfnm{Douglas}\binits{D.}} \AND
\bauthor{\bsnm{Maechler},~\bfnm{Martin}\binits{M.}}
(\byear{2010}).
\bhowpublished{lme4: Linear mixed-effects models using S4 classes.
R~package version 0.999375-34.}
\end{bmisc}
%
\bptok{imsref}%
\endbibitem

\bibitem[\protect\citeauthoryear{Breslow et~al.}{2009a}]{pmid20174455}
\begin{barticle}[pbm]
\bauthor{\bsnm{Breslow},~\bfnm{Norman~E.}\binits{N.~E.}},
\bauthor{\bsnm{Lumley},~\bfnm{Thomas}\binits{T.}},
\bauthor{\bsnm{Ballantyne},~\bfnm{Christie~M.}\binits{C.~M.}},
\bauthor{\bsnm{Chambless},~\bfnm{Lloyd~E.}\binits{L.~E.}} \AND
\bauthor{\bsnm{Kulich},~\bfnm{Michal}\binits{M.}}
(\byear{2009}a).
\btitle{Improved Horvitz--Thompson estimation of model parameters from two-phase stratified samples: Applications in epidemiology}.
\bjournal{Stat. Biosci.}
\bvolume{1}
\bpages{32}.
\bid{doi={10.1007/s12561-009-9001-6}, issn={1867-1764}, mid={NIHMS137324}, pmcid={2822363}, pmid={20174455}}
\end{barticle}
%
\bptok{imsref}%
\endbibitem

\bibitem[\protect\citeauthoryear{Breslow et~al.}{2009b}]{pmid19357328}
\begin{barticle}[pbm]
\bauthor{\bsnm{Breslow},~\bfnm{Norman~E.}\binits{N.~E.}},
\bauthor{\bsnm{Lumley},~\bfnm{Thomas}\binits{T.}},
\bauthor{\bsnm{Ballantyne},~\bfnm{Christie~M.}\binits{C.~M.}},
\bauthor{\bsnm{Chambless},~\bfnm{Lloyd~E.}\binits{L.~E.}} \AND
\bauthor{\bsnm{Kulich},~\bfnm{Michal}\binits{M.}}
(\byear{2009}b).
\btitle{Using the whole cohort in the analysis of case-cohort data}.
\bjournal{Am. J. Epidemiol.}
\bvolume{169}
\bpages{1398--1405}.
\bid{doi={10.1093/aje/kwp055}, issn={1476-6256}, pii={kwp055}, pmcid={2768499}, pmid={19357328}}
\end{barticle}
%
\bptok{imsref}%
\endbibitem

\bibitem[\protect\citeauthoryear{B$\stackrel{\fontsize{5pt}{7pt}\selectfont\mbox{$\circ$}}{\mbox{u}}${\v{z}}kov{\'a} and Lumley}{2009}]{pmid19035546}
\begin{barticle}[mr]
\bauthor{\bsnm{B$\stackrel{\fontsize{5pt}{7pt}\selectfont\mbox{$\circ$}}{\mbox{u}}${\v{z}}kov{\'a}},~\bfnm{Petra}\binits{P.}} \AND
\bauthor{\bsnm{Lumley},~\bfnm{Thomas}\binits{T.}}
(\byear{2009}).
\btitle{Semiparametric modeling of repeated measurements under outcome-dependent follow-up}.
\bjournal{Stat. Med.}
\bvolume{28}
\bpages{987--1003}.
\bid{doi={10.1002/sim.3496}, issn={0277-6715}, mr={2518361}}
\end{barticle}
%
\bptok{imsref}%
\endbibitem


\bibitem[\protect\citeauthoryear{Carroll et~al.}{2006}]{carroll2006measurement}
\begin{bbook}[mr]
\bauthor{\bsnm{Carroll},~\bfnm{Raymond~J.}\binits{R.~J.}},
\bauthor{\bsnm{Ruppert},~\bfnm{David}\binits{D.}},
\bauthor{\bsnm{Stefanski},~\bfnm{Leonard~A.}\binits{L.~A.}} \AND
\bauthor{\bsnm{Crainiceanu},~\bfnm{Ciprian~M.}\binits{C.~M.}}
(\byear{2006}).
\btitle{Measurement Error in Nonlinear Models: A Modern Perspective},
\bedition{2nd} ed.
\bseries{Monographs on Statistics and Applied Probability}
\bvolume{105}.
\bpublisher{Chapman \& Hall/CRC},
\blocation{Boca Raton, FL}.
\bid{doi={10.1201/9781420010138}, mr={2243417}}
\end{bbook}
%
\bptok{imsref}%
\endbibitem

\bibitem[\protect\citeauthoryear{Fisher}{1936}]{Fisher1936}
\begin{barticle}[author]
\bauthor{\bsnm{Fisher},~\bfnm{R.~A.}\binits{R.~A.}}
(\byear{1936}).
\btitle{The use of multiple measurements in taxonomic problems}.
\bjournal{Annals of Eugenics}
\bvolume{7}
\bpages{179--188}.
\end{barticle}
%
\bptok{imsref}%
\endbibitem

\bibitem[\protect\citeauthoryear{CAMP Research Group}{1999}]{pmid10027502}
\begin{barticle}[pbm]
\bauthor{\bsnm{CAMP Research Group}}
(\byear{1999}).
\btitle{The childhood asthma management program (CAMP): Design, rationale, and methods. Childhood asthma management program research group}.
\bjournal{Control. Clin. Trials}
\bvolume{20}
\bpages{91--120}.
\bid{issn={0197-2456}, pii={S0197245698000440}, pmid={10027502}}
\end{barticle}
%
\bptok{imsref}%
\endbibitem

\bibitem[\protect\citeauthoryear{CAMP Research Group}{2000}]{CAMP2000}
\begin{barticle}[author]
\bauthor{\bsnm{CAMP Research Group}}
(\byear{2000}).
\btitle{Long-term effects of budesonide or nedocrimil in children with asthma}.
\bjournal{N. Engl. J. Med.}
\bvolume{343}
\bpages{1054--1063}.
\end{barticle}
%
\bptok{imsref}%
\endbibitem

\bibitem[\protect\citeauthoryear{Horvitz and Thompson}{1952}]{H52}
\begin{barticle}[mr]
\bauthor{\bsnm{Horvitz},~\bfnm{D.~G.}\binits{D.~G.}} \AND
\bauthor{\bsnm{Thompson},~\bfnm{D.~J.}\binits{D.~J.}}
(\byear{1952}).
\btitle{A generalization of sampling without replacement from a finite universe}.
\bjournal{J. Amer. Statist. Assoc.}
\bvolume{47}
\bpages{663--685}.
\bid{issn={0162-1459}, mr={0053460}}
\end{barticle}
%
\bptok{imsref}%
\endbibitem


\bibitem[\protect\citeauthoryear{Kish}{1965}]{kish1965survey}
\begin{bbook}[author]
\bauthor{\bsnm{Kish},~\bfnm{Leslie}\binits{L.}}
(\byear{1965}).
\btitle{Survey Sampling}.
\bpublisher{Wiley},
\blocation{New York}.
\end{bbook}
%
\bptok{imsref}%
\endbibitem

\bibitem[\protect\citeauthoryear{Korn and Graubard}{2011}]{korn2011analysis}
\begin{bbook}[author]
\bauthor{\bsnm{Korn},~\bfnm{Edward~L.}\binits{E.~L.}} \AND
\bauthor{\bsnm{Graubard},~\bfnm{Barry~I.}\binits{B.~I.}}
(\byear{2011}).
\btitle{Analysis of Health Surveys}.
\bpublisher{Wiley},
\blocation{New York}.
\end{bbook}
%
\bptok{imsref}%
\endbibitem

\bibitem[\protect\citeauthoryear{Laird and Ware}{1982}]{pmid7168798}
\begin{barticle}[pbm]
\bauthor{\bsnm{Laird},~\bfnm{N.~M.}\binits{N.~M.}} \AND
\bauthor{\bsnm{Ware},~\bfnm{J.~H.}\binits{J.~H.}}
(\byear{1982}).
\btitle{Random-effects models for longitudinal data}.
\bjournal{Biometrics}
\bvolume{38}
\bpages{963--974}.
\bid{issn={0006-341X}, pmid={7168798}}
\end{barticle}
%
\bptok{imsref}%
\endbibitem

\bibitem[\protect\citeauthoryear{Lawless, Kalbfleisch and Wild}{1999}]{ISI000079074800009}
\begin{barticle}[mr]
\bauthor{\bsnm{Lawless},~\bfnm{J.~F.}\binits{J.~F.}},
\bauthor{\bsnm{Kalbfleisch},~\bfnm{J.~D.}\binits{J.~D.}} \AND
\bauthor{\bsnm{Wild},~\bfnm{C.~J.}\binits{C.~J.}}
(\byear{1999}).
\btitle{Semiparametric methods for response-selective and missing data problems in regression}.
\bjournal{J. R. Stat. Soc. Ser. B. Stat. Methodol.}
\bvolume{61}
\bpages{413--438}.
\bid{doi={10.1111/1467-9868.00185}, issn={1369-7412}, mr={1680310}}
\end{barticle}
%
\bptok{imsref}%
\endbibitem

\bibitem[\protect\citeauthoryear{Lin and Ying}{2001}]{ISI000167326000010}
\begin{barticle}[mr]
\bauthor{\bsnm{Lin},~\bfnm{D.~Y.}\binits{D.~Y.}} \AND
\bauthor{\bsnm{Ying},~\bfnm{Z.}\binits{Z.}}
(\byear{2001}).
\btitle{Semiparametric and nonparametric regression analysis of longitudinal data}.
\bjournal{J. Amer. Statist. Assoc.}
\bvolume{96}
\bpages{103--126}.
\bid{doi={10.1198/016214501750333018}, issn={0162-1459}, mr={1952726}}
\bptnote{check related, check pages}%
\end{barticle}
%
\bptok{imsref}%
\endbibitem

\bibitem[\protect\citeauthoryear{Lipsitz et~al.}{2002}]{pmid12229997}
\begin{barticle}[mr]
\bauthor{\bsnm{Lipsitz},~\bfnm{Stuart~R.}\binits{S.~R.}},
\bauthor{\bsnm{Fitzmaurice},~\bfnm{Garrett~M.}\binits{G.~M.}},
\bauthor{\bsnm{Ibrahim},~\bfnm{Joseph~G.}\binits{J.~G.}},
\bauthor{\bsnm{Gelber},~\bfnm{Richard}\binits{R.}} \AND
\bauthor{\bsnm{Lipshultz},~\bfnm{Steven}\binits{S.}}
(\byear{2002}).
\btitle{Parameter estimation in longitudinal studies with outcome-dependent follow-up}.
\bjournal{Biometrics}
\bvolume{58}
\bpages{621--630}.
\bid{doi={10.1111/j.0006-341X.2002.00621.x}, issn={0006-341X}, mr={1933535}}
\end{barticle}
%
\bptok{imsref}%
\endbibitem


\bibitem[\protect\citeauthoryear{Little and Rubin}{2002}]{LittleRubin2002}
\begin{bbook}[mr]
\bauthor{\bsnm{Little},~\bfnm{Roderick~J.~A.}\binits{R.~J.~A.}} \AND
\bauthor{\bsnm{Rubin},~\bfnm{Donald~B.}\binits{D.~B.}}
(\byear{2002}).
\btitle{Statistical Analysis with Missing Data},
\bedition{2nd} ed.
\bpublisher{Wiley},
\blocation{Hoboken, NJ}.
\bid{doi={10.1002/9781119013563}, mr={1925014}}
\end{bbook}
%
\bptok{imsref}%
\endbibitem

\bibitem[\protect\citeauthoryear{Little and Schluchter}{1985}]{little1985maximum}
\begin{barticle}[mr]
\bauthor{\bsnm{Little},~\bfnm{Roderick~J.~A.}\binits{R.~J.~A.}} \AND
\bauthor{\bsnm{Schluchter},~\bfnm{Mark~D.}\binits{M.~D.}}
(\byear{1985}).
\btitle{Maximum likelihood estimation for mixed continuous and categorical data with missing values}.
\bjournal{Biometrika}
\bvolume{72}
\bpages{497--512}.
\bid{doi={10.1093/biomet/72.3.497}, issn={0006-3444}, mr={0817564}}
\end{barticle}
%
\bptok{imsref}%
\endbibitem

\bibitem[\protect\citeauthoryear{Lyon et al.}{2004}]{L04}
\begin{barticle}[pbm]
\bauthor{\bsnm{Lyon},~\bfnm{Helen}\binits{H.}},
\bauthor{\bsnm{Lange},~\bfnm{Christoph}\binits{C.}},
\bauthor{\bsnm{Lake},~\bfnm{Stephen}\binits{S.}},
\bauthor{\bsnm{Silverman},~\bfnm{Edwin~K.}\binits{E.~K.}},
\bauthor{\bsnm{Randolph},~\bfnm{Adrienne~G.}\binits{A.~G.}},
\bauthor{\bsnm{Kwiatkowski},~\bfnm{David}\binits{D.}},
\bauthor{\bsnm{Raby},~\bfnm{Benjamin~A.}\binits{B.~A.}},
\bauthor{\bsnm{Lazarus},~\bfnm{Ross}\binits{R.}},
\bauthor{\bsnm{Weiland},~\bfnm{Katy~M.}\binits{K.~M.}},
\bauthor{\bsnm{Laird},~\bfnm{Nan}\binits{N.}} \AND
\bauthor{\bsnm{Weiss},~\bfnm{Scott~T.}\binits{S.~T.}}
(\byear{2004}).
\btitle{IL10 gene polymorphisms are associated with asthma phenotypes in children}.
\bjournal{Genet. Epidemiol.}
\bvolume{26}
\bpages{155--165}.
\bid{doi={10.1002/gepi.10298}, issn={0741-0395}, mid={NIHMS389827}, pmcid={3705717}, pmid={14748015}}
\end{barticle}
%
\bptok{imsref}%
\endbibitem

\bibitem[\protect\citeauthoryear{Marti and Chavance}{2011}]{pmid21351290}
\begin{barticle}[mr]
\bauthor{\bsnm{Marti},~\bfnm{Helena}\binits{H.}} \AND
\bauthor{\bsnm{Chavance},~\bfnm{Michel}\binits{M.}}
(\byear{2011}).
\btitle{Multiple imputation analysis of case-cohort studies}.
\bjournal{Stat. Med.}
\bvolume{30}
\bpages{1595--1607}.
\bid{doi={10.1002/sim.4130}, issn={0277-6715}, mr={2828892}}
\end{barticle}
%
\bptok{imsref}%
\endbibitem

\bibitem[\protect\citeauthoryear{Neuhaus, Scott and Wild}{2002}]{NeuhScotWildanal2002}
\begin{barticle}[mr]
\bauthor{\bsnm{Neuhaus},~\bfnm{J.}\binits{J.}},
\bauthor{\bsnm{Scott},~\bfnm{A.~J.}\binits{A.~J.}} \AND
\bauthor{\bsnm{Wild},~\bfnm{C.~J.}\binits{C.~J.}}
(\byear{2002}).
\btitle{The analysis of retrospective family studies}.
\bjournal{Biometrika}
\bvolume{89}
\bpages{23--37}.\
\bid{doi={10.1093/biomet/89.1.23}, issn={0006-3444}, mr={1888343}}
\end{barticle}
%
\bptok{imsref}%
\endbibitem

\bibitem[\protect\citeauthoryear{Neuhaus, Scott and Wild}{2006}]{NeuhScotWildfami2006}
\begin{barticle}[mr]
\bauthor{\bsnm{Neuhaus},~\bfnm{J.~M.}\binits{J.~M.}},
\bauthor{\bsnm{Scott},~\bfnm{A.~J.}\binits{A.~J.}} \AND
\bauthor{\bsnm{Wild},~\bfnm{C.~J.}\binits{C.~J.}}
(\byear{2006}).
\btitle{Family-specific approaches to the analysis of case--control family data}.
\bjournal{Biometrics}
\bvolume{62}
\bpages{488--494}.
\bid{doi={10.1111/j.1541-0420.2005.00450.x}, issn={0006-341X}, mr={2236831}}
\end{barticle}
%
\bptok{imsref}%
\endbibitem

\bibitem[\protect\citeauthoryear{Neuhaus et~al.}{2014}]{pmid24571396}
\begin{barticle}[mr]
\bauthor{\bsnm{Neuhaus},~\bfnm{John~M.}\binits{J.~M.}},
\bauthor{\bsnm{Scott},~\bfnm{Alastair~J.}\binits{A.~J.}},
\bauthor{\bsnm{Wild},~\bfnm{Christopher~J.}\binits{C.~J.}},
\bauthor{\bsnm{Jiang},~\bfnm{Yannan}\binits{Y.}},
\bauthor{\bsnm{McCulloch},~\bfnm{Charles~E.}\binits{C.~E.}} \AND
\bauthor{\bsnm{Boylan},~\bfnm{Ross}\binits{R.}}
(\byear{2014}).
\btitle{Likelihood-based analysis of longitudinal data from outcome-related sampling designs}.
\bjournal{Biometrics}
\bvolume{70}
\bpages{44--52}.
\bid{doi={10.1111/biom.12108}, issn={0006-341X}, mr={3251665}}
\end{barticle}
%
\bptok{imsref}%
\endbibitem

\bibitem[\protect\citeauthoryear{R Core Team}{2013}]{R}
\begin{bmisc}[author]
\borganization{R Core Team}
(\byear{2013}).
\bhowpublished{\textit{R: A Language and Environment for Statistical Computing}.
R Foundation for Statistical Computing,
Vienna, Austria.}
\end{bmisc}
%
\bptok{imsref}%
\endbibitem

\bibitem[\protect\citeauthoryear{Raghunathan et~al.}{2001}]{Raghunathan++01}
\begin{barticle}[author]
\bauthor{\bsnm{Raghunathan},~\bfnm{Trivellore~E.}\binits{T.~E.}},
\bauthor{\bsnm{Lepkowski},~\bfnm{James~M.}\binits{J.~M.}},
\bauthor{\bsnm{Hoewyk},~\bfnm{John~V.}\binits{J.~V.}} \AND
\bauthor{\bsnm{Solenberger},~\bfnm{Peter}\binits{P.}}
(\byear{2001}).
\btitle{A multivariate technique for multiply imputing missing values using a sequence of regression models}.
\bjournal{Surv. Methodol.}
\bvolume{27}
\bpages{85--95}.
\end{barticle}
%
\bptok{imsref}%
\endbibitem

\bibitem[\protect\citeauthoryear{Robins, Rotnitzky and Zhao}{1994}]{RobiRotnZhaoesti1994}
\begin{barticle}[mr]
\bauthor{\bsnm{Robins},~\bfnm{James~M.}\binits{J.~M.}},
\bauthor{\bsnm{Rotnitzky},~\bfnm{Andrea}\binits{A.}} \AND
\bauthor{\bsnm{Zhao},~\bfnm{Lue~Ping}\binits{L.~P.}}
(\byear{1994}).
\btitle{Estimation of regression coefficients when some regressors are not always observed}.
\bjournal{J. Amer. Statist. Assoc.}
\bvolume{89}
\bpages{846--866}.
\bid{issn={0162-1459}, mr={1294730}}
\end{barticle}
%
\bptok{imsref}%
\endbibitem

\bibitem[\protect\citeauthoryear{Rubin}{1976}]{ISIA1976CP66700021}
\begin{barticle}[mr]
\bauthor{\bsnm{Rubin},~\bfnm{Donald~B.}\binits{D.~B.}}
(\byear{1976}).
\btitle{Inference and missing data}.
\bjournal{Biometrika}
\bvolume{63}
\bpages{581--592}.
\bid{issn={0006-3444}, mr={0455196}}
\bptnote{check related, check pages}%
\end{barticle}
%
\bptok{imsref}%
\endbibitem

\bibitem[\protect\citeauthoryear{Schafer}{2010}]{schafer2010analysis}
\begin{bbook}[author]
\bauthor{\bsnm{Schafer},~\bfnm{Joseph~L.}\binits{J.~L.}}
(\byear{2010}).
\btitle{Analysis of Incomplete Multivariate Data}.
\bpublisher{CRC Press},
\blocation{Boca Raton, FL}.
\end{bbook}
%
\bptok{imsref}%
\endbibitem

\bibitem[\protect\citeauthoryear{Schafer and Graham}{2002}]{SchaferGraham2002}
\begin{barticle}[pbm]
\bauthor{\bsnm{Schafer},~\bfnm{Joseph~L.}\binits{J.~L.}} \AND
\bauthor{\bsnm{Graham},~\bfnm{John~W.}\binits{J.~W.}}
(\byear{2002}).
\btitle{Missing data: Our view of the state of the art}.
\bjournal{Psychol. Methods}
\bvolume{7}
\bpages{147--177}.
\bid{issn={1082-989X}, pmid={12090408}}
\end{barticle}
%
\bptok{imsref}%
\endbibitem

\bibitem[\protect\citeauthoryear{Schildcrout, Garbett and Heagerty}{2013}]{pmid23409789}
\begin{barticle}[mr]
\bauthor{\bsnm{Schildcrout},~\bfnm{Jonathan~S.}\binits{J.~S.}},
\bauthor{\bsnm{Garbett},~\bfnm{Shawn~P.}\binits{S.~P.}} \AND
\bauthor{\bsnm{Heagerty},~\bfnm{Patrick~J.}\binits{P.~J.}}
(\byear{2013}).
\btitle{Outcome vector dependent sampling with longitudinal continuous response data: Stratified sampling based on summary statistics}.
\bjournal{Biometrics}
\bvolume{69}
\bpages{405--416}.
\bid{doi={10.1111/biom.12013}, issn={0006-341X}, mr={3071059}}
\end{barticle}
%
\bptok{imsref}%
\endbibitem

\bibitem[\protect\citeauthoryear{Schildcrout and Heagerty}{2008}]{pmid18372397}
\begin{barticle}[pbm]
\bauthor{\bsnm{Schildcrout},~\bfnm{Jonathan~S.}\binits{J.~S.}} \AND
\bauthor{\bsnm{Heagerty},~\bfnm{Patrick~J.}\binits{P.~J.}}
(\byear{2008}).
\btitle{On outcome-dependent sampling designs for longitudinal binary response data with time-varying covariates}.
\bjournal{Biostatistics}
\bvolume{9}
\bpages{735--749}.
\bid{doi={10.1093/biostatistics/kxn006}, issn={1468-4357}, pii={kxn006}, pmcid={2733177}, pmid={18372397}}
\end{barticle}
%
\bptok{imsref}%
\endbibitem

\bibitem[\protect\citeauthoryear{Schildcrout and Heagerty}{2011}]{pmid21457191}
\begin{barticle}[mr]
\bauthor{\bsnm{Schildcrout},~\bfnm{Jonathan~S.}\binits{J.~S.}} \AND
\bauthor{\bsnm{Heagerty},~\bfnm{Patrick~J.}\binits{P.~J.}}
(\byear{2011}).
\btitle{Outcome-dependent sampling from existing cohorts with longitudinal binary response data: Study planning and analysis}.
\bjournal{Biometrics}
\bvolume{67}
\bpages{1583--1593}.
\bid{doi={10.1111/j.1541-0420.2011.01582.x}, issn={0006-341X}, mr={2872409}}
\end{barticle}
%
\bptok{imsref}%
\endbibitem

\bibitem[\protect\citeauthoryear{Schildcrout and Rathouz}{2010}]{pmid19673861}
\begin{barticle}[mr]
\bauthor{\bsnm{Schildcrout},~\bfnm{Jonathan~S.}\binits{J.~S.}} \AND
\bauthor{\bsnm{Rathouz},~\bfnm{Paul~J.}\binits{P.~J.}}
(\byear{2010}).
\btitle{Longitudinal studies of binary response data following case--control and stratified case--control sampling: Design and analysis}.
\bjournal{Biometrics}
\bvolume{66}
\bpages{365--373}.
\bid{doi={10.1111/j.1541-0420.2009.01306.x}, issn={0006-341X}, mr={2758816}}
\end{barticle}
%
\bptok{imsref}%
\endbibitem

\bibitem[\protect\citeauthoryear{Schildcrout et~al.}{2012}]{pmid22086716}
\begin{barticle}[mr]
\bauthor{\bsnm{Schildcrout},~\bfnm{Jonathan~S.}\binits{J.~S.}},
\bauthor{\bsnm{Mumford},~\bfnm{Sunni~L.}\binits{S.~L.}},
\bauthor{\bsnm{Chen},~\bfnm{Zhen}\binits{Z.}},
\bauthor{\bsnm{Heagerty},~\bfnm{Patrick~J.}\binits{P.~J.}} \AND
\bauthor{\bsnm{Rathouz},~\bfnm{Paul~J.}\binits{P.~J.}}
(\byear{2012}).
\btitle{Outcome-dependent sampling for longitudinal binary response data based on a time-varying auxiliary variable}.
\bjournal{Stat. Med.}
\bvolume{31}
\bpages{2441--2456}.
\bid{doi={10.1002/sim.4359}, issn={0277-6715}, mr={2972258}}
\end{barticle}
%
\bptok{imsref}%
\endbibitem

\bibitem[\protect\citeauthoryear{Schildcrout et al.}{2015}]{supp}
\begin{bmisc}[author]
{\bauthor{\bsnm{Schildcrout},~\binits{J. S.}},
\bauthor{\bsnm{Rathouz},~\binits{P. J.}},
\bauthor{\bsnm{Zelnick},~\binits{L. R.}},
\bauthor{\bsnm{Garbett},~\binits{S. P.}} \AND
\bauthor{\bsnm{Heagerty},~\binits{P. J.}}}
(\byear{2015}).
\bhowpublished{Supplement to ``Biased sampling designs to improve research efficiency: Factors
influencing pulmonary function over time in children with
asthma.'' DOI:\doiurl{10.1214/15-AOAS826SUPPA}, DOI:\doiurl{10.1214/15-AOAS826SUPPB}}.
\bptok{imsref}%
\end{bmisc}
\endbibitem

\bibitem[\protect\citeauthoryear{Van~Buuren}{2012}]{van2012flexible}
\begin{bbook}[author]
\bauthor{\bsnm{Van~Buuren},~\bfnm{Stef}\binits{S.}}
(\byear{2012}).
\btitle{Flexible Imputation of Missing Data}.
\bpublisher{CRC Press},
\blocation{Boca Raton, FL}.
\end{bbook}
%
\bptok{imsref}%
\endbibitem

\bibitem[\protect\citeauthoryear{Weaver and Zhou}{2005}]{WeavZhouan2005}
\begin{barticle}[mr]
\bauthor{\bsnm{Weaver},~\bfnm{Mark~A.}\binits{M.~A.}} \AND
\bauthor{\bsnm{Zhou},~\bfnm{Haibo}\binits{H.}}
(\byear{2005}).
\btitle{An estimated likelihood method for continuous outcome regression models with outcome-dependent sampling}.
\bjournal{J. Amer. Statist. Assoc.}
\bvolume{100}
\bpages{459--469}.
\bid{doi={10.1198/016214504000001853}, issn={0162-1459}, mr={2160550}}
\end{barticle}
%
\bptok{imsref}%
\endbibitem

\bibitem[\protect\citeauthoryear{White, Royston and Wood}{2011}]{ISI000287106200008}
\begin{barticle}[mr]
\bauthor{\bsnm{White},~\bfnm{Ian~R.}\binits{I.~R.}},
\bauthor{\bsnm{Royston},~\bfnm{Patrick}\binits{P.}} \AND
\bauthor{\bsnm{Wood},~\bfnm{Angela~M.}\binits{A.~M.}}
(\byear{2011}).
\btitle{Multiple imputation using chained equations: Issues and guidance for practice}.
\bjournal{Stat. Med.}
\bvolume{30}
\bpages{377--399}.
\bid{doi={10.1002/sim.4067}, issn={0277-6715}, mr={2758870}}
\bptnote{check pages}%
\end{barticle}
%
\bptok{imsref}%
\endbibitem

\bibitem[\protect\citeauthoryear{Zhou et~al.}{2002}]{ZhouWeavQinLongWangsemi2002}
\begin{barticle}[mr]
\bauthor{\bsnm{Zhou},~\bfnm{Haibo}\binits{H.}},
\bauthor{\bsnm{Weaver},~\bfnm{M.~A.}\binits{M.~A.}},
\bauthor{\bsnm{Qin},~\bfnm{J.}\binits{J.}},
\bauthor{\bsnm{Longnecker},~\bfnm{M.~P.}\binits{M.~P.}} \AND
\bauthor{\bsnm{Wang},~\bfnm{M.~C.}\binits{M.~C.}}
(\byear{2002}).
\btitle{A semiparametric empirical likelihood method for data from an outcome-dependent sampling scheme with a continuous outcome}.
\bjournal{Biometrics}
\bvolume{58}
\bpages{413--421}.
\bid{doi={10.1111/j.0006-341X.2002.00413.x}, issn={0006-341X}, mr={1908182}}
\end{barticle}
%
\bptok{imsref}%
\endbibitem

\bibitem[\protect\citeauthoryear{Zhou et~al.}{2007}]{pmid17568219}
\begin{barticle}[author]
\bauthor{\bsnm{Zhou},~\bfnm{H.}\binits{H.}},
\bauthor{\bsnm{Chen},~\bfnm{J.}\binits{J.}},
\bauthor{\bsnm{Rissanen},~\bfnm{T.~H.}\binits{T.~H.}},
\bauthor{\bsnm{Korrick},~\bfnm{S.~A.}\binits{S.~A.}},
\bauthor{\bsnm{Hu},~\bfnm{H.}\binits{H.}},
\bauthor{\bsnm{Salonen},~\bfnm{J.~T.}\binits{J.~T.}} \AND
\bauthor{\bsnm{Longnecker},~\bfnm{M.~P.}\binits{M.~P.}}
(\byear{2007}).
\btitle{Outcome-dependent sampling: An efficient sampling and inference procedure for studies with a continuous outcome}.
\bjournal{Epidemiology}
\bvolume{18}\
\bpages{461--468}.
\end{barticle}
%
\bptok{imsref}%
\endbibitem

\bibitem[\protect\citeauthoryear{Zhou et~al.}{2011}]{pmid21252082}
\begin{barticle}[pbm]
\bauthor{\bsnm{Zhou},~\bfnm{Haibo}\binits{H.}},
\bauthor{\bsnm{Wu},~\bfnm{Yuanshan}\binits{Y.}},
\bauthor{\bsnm{Liu},~\bfnm{Yanyan}\binits{Y.}} \AND
\bauthor{\bsnm{Cai},~\bfnm{Jianwen}\binits{J.}}
(\byear{2011}).
\btitle{Semiparametric inference for a 2-stage outcome-auxiliary-dependent sampling design with continuous outcome}.
\bjournal{Biostatistics}
\bvolume{12}
\bpages{521--534}.
\bid{doi={10.1093/biostatistics/kxq080}, issn={1468-4357}, pii={kxq080}, pmcid={3114654}, pmid={21252082}}
\end{barticle}
%
\bptok{imsref}%
\endbibitem

\end{thebibliography}
\end{document}